\documentclass[journal]{IEEEtran}

\IEEEoverridecommandlockouts
\usepackage{cite}
\usepackage{graphicx}

\usepackage{subcaption}
\usepackage{textcomp}
\usepackage{amsmath}
\usepackage{amsfonts}
\usepackage{array}
\usepackage{multirow}
\usepackage{tabulary}
\usepackage{algpseudocode}
\usepackage{bm}
\usepackage{float}
\usepackage{stfloats}
\usepackage[linesnumbered,ruled,vlined]{algorithm2e}
\usepackage{amssymb} % để dùng ▷

\usepackage{booktabs}
\usepackage{hyperref} % To be commented once finished
\usepackage[table]{xcolor}
\usepackage{collcell}
\usepackage{hhline}
\usepackage{pgf}
\usepackage{multirow}

\def\colorModel{hsb} %You can use rgb or hsb

\newcommand\ColCell[1]{
  \pgfmathparse{#1<50?1:0}  %Threshold for changing the font color into the cells
    \ifnum\pgfmathresult=0\relax\color{white}\fi
    \pgfmathsetmacro\compA{0}      %Component R or H
    \pgfmathsetmacro\compB{#1/100} %Component G or S
    \pgfmathsetmacro\compC{1}      %Component B or B
  \edef\x{\noexpand\centering\noexpand\cellcolor[\colorModel]{\compA,\compB,\compC}}\x #1
  } 
\newcolumntype{E}{>{\collectcell\ColCell}m{0.4cm}<{\endcollectcell}}  %Cell width

\usepackage{adjustbox} % For scaling

\begin{document}

\markboth{Preprint}%
{Nguyen \MakeLowercase{\textit{et al.}}: Secure and Efficient UAV Face Detection Leveraging Homomorphic Encryption and Mobile Edge Computing}

\title{Secure and Efficient UAV-Based Face Detection via Homomorphic Encryption and Edge Computing}

\makeatletter
\newcommand{\linebreakand}{%
\end{@IEEEauthorhalign}
\hfill\mbox{}\par
\mbox{}\hfill\begin{@IEEEauthorhalign}
}
\makeatother
	
\author{
    Nguyen Van Duc, 
    Bui Duc Manh, 
    Quang-Trung Luu,
    Dinh Thai Hoang, \\
    Van-Linh Nguyen,
    and 
    Diep N. Nguyen%
    \thanks{
        Nguyen Van Duc and Quang-Trung~Luu
        are with School of Electrical and Electronic Engineering, Hanoi University of Science and Technology, Hanoi 100000, Vietnam (emails: \{duc.nv213886@sis., trung.luuquang@\}hust.edu.vn).}%
    \thanks{
        Bui Duc Manh (work done while at UTS), Dinh Thai Hoang, and Diep N.~Nguyen are with School of Electrical and Data Engineering, University of Technology Sydney, Australia (emails: \{ducmanh.bui, hoang.dinh, diep.nguyen\}@uts.edu.au).}%
    \thanks{
    Van-Linh Nguyen is with the Department of Computer Science and Information Engineering, National Chung Cheng University, Taiwan (email: nvlinh@cs.ccu.edu.tw).}%
}

\maketitle

\begin{abstract}
%Đức viết
This paper aims to propose a novel machine learning (ML) approach incorporating Homomorphic Encryption (HE) to address privacy limitations in Unmanned Aerial Vehicles (UAV)-based face detection. Due to challenges related to distance, altitude, and face orientation, high-resolution imagery and sophisticated neural networks enable accurate face recognition in dynamic environments. However, privacy concerns arise from the extensive surveillance capabilities of UAVs. To resolve this issue, we propose a novel framework that integrates HE with advanced neural networks to secure facial data throughout the inference phase. This method ensures that facial data remains secure with minimal impact on detection accuracy. Specifically, the proposed system leverages the  Cheon–Kim–Kim–Song (CKKS) scheme to perform computations directly on encrypted data, optimizing computational efficiency and security. 
Furthermore, we develop an effective data encoding method specifically designed to preprocess the raw facial data into CKKS form in a Single-Instruction-Multiple-Data (SIMD) manner. Building on this, we design a secure inference algorithm to compute on ciphertext without needing decryption. This approach not only protects data privacy during the processing of facial data but also enhances the efficiency of UAV-based face detection systems. 
% Furthermore, we develop an effective training algorithm specifically designed to handle the challenges of processing encrypted data. This algorithm allows the system to compute directly on encrypted ciphertexts without needing decryption. This approach not only protects data privacy during the processing of facial data but also enhances the efficiency of UAV-based face detection systems. 
Experimental results demonstrate that our method effectively balances privacy protection and detection performance, making it a viable solution for UAV-based secure face detection. Significantly, our approach (while maintaining data confidentially with HE encryption) can still achieve an accuracy of less than 1\% compared to the benchmark without using encryption. 
\end{abstract}

\begin{IEEEkeywords}
UAVs, face detection, homomorphic encryption, privacy-preserving, deep learning, mobile edge computing.
\end{IEEEkeywords}

%===============================================
%===============================================
\section{Introduction}
%Đứcviết
%Para 1: Topic (UAV/Drone-based face detection): Write one paragraph to introduce this topic

%---------------------------------
%---------------------------------
\subsection{Motivations}

\IEEEPARstart{T}{he}
development of Unmanned Aerial Vehicles (UAVs), commonly known as \emph{drones}, has revolutionized various fields by combining advanced mobility, autonomy, and sensing capabilities~\cite{javaid2023communication,sahoo2022toward}. Equipped with high-resolution cameras and sophisticated onboard processing units, UAVs are increasingly used for face detection, an application that has significant implications in areas such as security, surveillance, disaster management, and crowd monitoring~\cite{wieczorek2021lightweight,diez2024efficient,peng2023skynet}. For instance, in the context of natural disasters such as flooding, UAVs can be deployed to perform aerial surveillance over affected regions, leveraging facial recognition algorithms to detect and identify stranded survivors, thereby facilitating timely and targeted rescue operations. In the domain of public safety, UAVs are also utilized to assist law enforcement agencies in identifying individuals involved in criminal activities or detecting potential threats in densely populated areas. Importantly, these applications are designed to integrate privacy-preserving mechanisms, such as encrypted data transmission~\cite{homomorphicEncryption2025} and anonymized facial feature extraction~\cite{privacyAdv2023}, to protect user privacy while maintaining operational effectiveness.. UAVs offer unparalleled advantages in face-detection tasks due to their ability to navigate challenging terrains, reach inaccessible or hazardous areas, and provide real-time data from dynamic perspectives. Their aerial vantage point enables wide-area coverage, making them efficient in monitoring large crowds or securing expansive perimeters. Furthermore, UAVs can be deployed rapidly, providing a cost-effective and scalable solution compared to fixed surveillance systems. These benefits make UAVs invaluable tools for addressing practical challenges in face detection, such as varying angles, lighting conditions, and complex environments, which are often encountered in real-world scenarios.

%Para 2: Write one paragraph to explain challenges when deploying face detection tasks on UAVs

Performing face detection tasks on UAVs presents significant technical challenges, primarily due to the limited battery life, computational power, and onboard storage of UAVs. Face detection is computationally intensive, requiring real-time image processing and deep learning (DL) inference~\cite{zhang2020refineface}, which places a heavy demand on UAV hardware. However, most UAVs are equipped with lightweight processors and limited memory, making it difficult to run complex deep learning models efficiently~\cite{mcenroe2022survey,heidari2023machine}. Moreover, UAVs have strict energy constraints, as continuous image processing significantly drains the battery, reducing flight time and operational effectiveness~\cite{mcenroe2022survey}. Additionally, real-time inference on UAVs is hindered by thermal limitations, where excessive heat from sustained processing can degrade system performance or lead to failures. As a result, many approaches have been proposed in the literature to address these constraints, including model compression, edge cloud collaboration, and hardware aware optimization techniques. However, balancing the trade off between face detection accuracy and the preservation of individual privacy remains a critical and underexplored challenge. Most existing solutions focus on improving computational efficiency or energy consumption, while relatively few consider how privacy preserving mechanisms such as homomorphic encryption or federated learning can be integrated without compromising detection performance.

% \edit{
% Beyond the direct application in UAV-based face detection, our proposed privacy-preserving inference framework can be seamlessly integrated into the emerging Open Radio Access Network (Open RAN) architecture. Open RAN decouples hardware and software components in cellular networks, enabling more flexible, interoperable, and intelligent network deployments \cite{brik2024explainable}. 
% In particular, the disaggregated and virtualized architecture of Open RAN allows AI-based tasks, e.g., facial recognition and anomaly detection, to be dynamically offloaded to edge nodes, while ensuring strict privacy requirements \cite{brik2024explainable,groen2024implementing}. Our approach, which enables secure deep learning inference via homomorphic encryption on encrypted facial data, is highly compatible with such an architecture. It empowers Open RAN nodes to process sensitive user data (e.g., identity or behavior) without decrypting it, enhancing both operational intelligence and data confidentiality.}

Recent research has also explored the deployment of UAVs within Open Radio Access Network (Open RAN) architectures to enhance communication, coordination, and security in collaborative missions~\cite{li2024energy}. However, this integration introduces new challenges related to privacy protection, as UAVs often collect sensitive personal data over open and potentially insecure wireless links~\cite{li2024energy,soleymani2024secure}. Our privacy-preserving inference framework addresses this challenge and is well-suited for integration with emerging Open RAN systems, enabling encrypted inference at edge nodes to ensure data confidentiality while supporting efficient task offloading.

% Recent research has also explored the deployment of UAVs within Open Radio Access Network (Open RAN) architectures to enhance communication, coordination, and security in collaborative missions \cite{li2024energy}. However, this integration introduces new challenges related to privacy protection, as UAVs often collect sensitive personal data over open and potentially insecure wireless links \cite{li2024energy,soleymani2024secure}. Our privacy-preserving inference framework addresses this challenge and is well-suited for integration with emerging Open RAN systems, enabling encrypted inference at edge nodes to ensure data confidentiality while supporting efficient task offloading.

%---------------------------------
%---------------------------------
\subsection{Related Work}

% \vspace{0.2cm}
\noindent{}\textbf{UAV-based face detection.}
Most of the current research works focus on optimizing machine learning models for face detection to make them work efficiently on resource-constrained UAVs. For example, in~\cite{luo2020lightweight}, the authors aim to address the challenge of achieving real-time, accurate face detection with lightweight CNNs by reducing computational complexity and model size. The key solution involves integrating the network with an image pyramid for efficient feature reuse, optimizing anchor selection, and applying a weight assignment strategy to exclude overly difficult training samples. As a result, the proposed detector achieves high accuracy on benchmarks like WIDERFACE and FDDB, runs at $50$ FPS on CPU, and maintains a compact model size under $0.1$M parameters, making it highly suitable for mobile and resource-constrained devices. In~\cite{wieczorek2021lightweight}, the authors introduce a lightweight convolutional neural network (CNN) architecture specifically designed for portable devices, targeting real-time face detection in challenging environments where faces may be partially obscured or blend into complex backgrounds. The core innovation lies in minimizing the number of pooling and convolution operations, significantly reducing computational overhead without sacrificing detection capability. Experimental results demonstrate that, despite its streamlined architecture with fewer layers, the proposed lightweight CNN achieves remarkable accuracy exceeding 99\% and precision above 98\% across various textured backgrounds. Similarly, several other studies, such as~\cite{ou2024aerialface, diez2024efficient}, have focused on optimizing CNN model parameters to meet the computational constraints of UAV platforms. These works strategically reduce model complexity while maintaining detection performance, ensuring compatibility with the limited processing power and energy budgets of UAVs. Experimental evaluations conducted on real UAV devices demonstrate their effectiveness in achieving an optimal trade-off among energy consumption, model size, and detection accuracy, highlighting their practicality for real-world UAV applications.

Although various approaches (e.g.,~\cite{luo2020lightweight,wieczorek2021lightweight,ou2024aerialface,diez2024efficient}) have been proposed to optimize ML models for efficient execution on UAV platforms, they encounter inherent and persistent limitations that are difficult to overcome. Specifically, most of these methods rely on techniques such as model compression, quantization, or pruning to reduce computational demands. While effective to some extent, such techniques inevitably compromise model accuracy, robustness, and generalization ability, particularly when deployed in complex, dynamic real-world environments. More importantly, despite these optimizations, UAVs’ fundamental hardware constraints, including limited processing power, memory capacity, and battery life, remain significant bottlenecks, making it highly challenging to achieve real-time, high-performance ML inference without adversely impacting flight duration or system stability. Additionally, the hardware and software ecosystems onboard UAVs often impose strict requirements, such as specific CPU/GPU capabilities, compatible operating systems, and pre-installed libraries that further restrict flexibility. Maintaining and updating ML models directly on UAVs also introduces operational difficulties, frequently requiring either physical access to devices or intricate remote update mechanisms, which increases system complexity and maintenance overhead. Consequently, there is a growing trend toward offloading ML-based face detection tasks to nearby mobile edge servers~\cite{mcenroe2022survey}, leveraging their superior computational resources, scalability, and ease of maintenance while allowing UAVs to focus on efficient data collection and transmission.

In particular, the authors in~\cite{koubaa2020deepbrain} proposed a new model for computation offloading in Internet-connected drones, aiming to address the limitations of resource-constrained UAVs. The core idea is to offload heavy, computation-intensive tasks, such as deep learning applications, to a nearby mobile edge device, thereby reducing the UAV’s energy consumption and extending its mission lifetime. This approach not only conserves onboard resources but also leverages the superior computational capabilities of edge servers to execute complex tasks more efficiently. Similarly, the authors in~\cite{peng2023skynet} considered an offloading strategy where intensive ML-based face detection tasks are delegated to edge devices for processing, demonstrating the feasibility and effectiveness of this paradigm. However, they addressed a more complex scenario involving collaborative multi-UAV operations for real-time personal identification and localization. In this system, each UAV captures images of the crowd and utilizes onboard lightweight human face detection models to extract face sub-images. These sub-images are then transmitted to a designated edge server, which performs joint processing by fusing images from multiple viewpoints to compute the 3D locations of individuals and generate identification results. This highlights the growing trend of utilizing edge computing to overcome the inherent limitations of UAV platforms while enabling advanced ML tasks in real-time. 

Despite offering significantly greater computational resources than UAVs, edge servers still encounter inherent limitations, such as constrained processing power, limited memory, and network variability in mobile or harsh environments, all of which can impact the real-time performance and scalability of face recognition systems assisted by edge computing. In contrast to cloud infrastructure, these hardware constraints necessitate the use of carefully optimized models. To address these challenges, our approach employs a lightweight CNN integrated with homomorphic encryption, deliberately chosen for its computational feasibility on resource-limited edge platforms. Although our current implementation is evaluated in a simulated environment, future work will involve deployment on actual edge platforms to rigorously assess performance under real-world constraints.

% Despite offering significantly greater computational resources than UAVs, edge servers still encounter inherent limitations, such as constrained processing power, limited memory, and network variability in mobile or harsh environments, all of which can impact the real-time performance and scalability of edge-assisted face recognition systems. In contrast to cloud infrastructure, these hardware constraints necessitate the use of carefully optimized models. To address these challenges, our approach employs a lightweight CNN integrated with homomorphic encryption, deliberately chosen for its computational feasibility on resource-limited edge platforms. Although our current implementation has been evaluated within a simulated environment, the potential impact of these constraints on system scalability and performance in real-world scenarios cannot be understated. As such, future work will focus on deploying the system on actual edge platforms to rigorously assess its performance under practical conditions.

\vspace{0.1cm}
\noindent{}\textbf{Privacy and security awareness.}
While leveraging edge devices offers clear advantages in efficiently handling ML-based face detection tasks, it simultaneously introduces significant privacy and security concerns~\cite{mcenroe2022survey}. Unlike other ML classification tasks, such as object detection or activity recognition, UAV-based face detection inherently raises more serious privacy risks~\cite{yao2024framework,lee2021development,IEEE_Facial_Recognition_Ethics}. UAVs equipped with high-resolution cameras are capable of collecting sensitive personal data from both public and private spaces, often from considerable distances and altitudes, typically without individuals’ knowledge or consent. This capability poses a substantial threat to privacy rights, as it enables the unauthorized capture of identifiable information. More critically, when the collected data, including human faces, location details, and surrounding environmental context, is transmitted from UAVs to edge servers for processing, it becomes highly susceptible to a variety of cyberattacks. Common threats include man-in-the-middle (MitM) attacks during wireless transmission, as well as malware injection and unauthorized access at the edge device level. The potential exploitation of such sensitive information, such as stealing, leaking, or faking identities, represents a severe and pressing challenge. To the best of our knowledge, existing literature does not adequately address the privacy-preserving issues specific to UAV-based face detection in a UAV-Edge communication architecture. Our work is the first to explicitly focus on this problem, offering a novel and practical solution to safeguard sensitive data throughout the entire UAV-to-edge processing pipeline.

\vspace{0.1cm}
\noindent{}\textbf{Homomorphic encryption for secure communications.}
In recent years, there has been growing concern about privacy and data security in communications between UAVs and edge servers, particularly with advances in optimizing deep learning models for resource-constrained UAVs~\cite{mao2023security}. In this context, homomorphic encryption (HE) has emerged as a promising solution, as it enables direct computation on encrypted data, ensuring privacy throughout transmission and processing. Although the theoretical foundations of HE have been known for decades, its realization in the form of fully homomorphic encryption schemes only became feasible after Craig Gentry’s groundbreaking work in 2009~\cite{gentry2009}. Since that key breakthrough, HE has evolved from a theoretical construct into a versatile tool deployed across various applications. One prominent application is in semantic communication systems, where it enables the processing and transmission of encrypted semantic information without requiring decryption, thereby enhancing data security in intelligent communication environments~\cite{meng2025}. In vehicular platooning control systems, HE has been employed to facilitate the secure exchange of sensitive information while preserving real-time performance~\cite{quero2023}. Despite these advancements, the implementation of HE in secure computation protocols for VANETs presents several challenges, including key management, noise budget, and computational overhead~\cite{sun_survey}. Addressing these concerns requires optimization strategies tailored to resource-constrained environments, where efficiency is crucial for practical deployment~\cite{adir2024}. Moreover, HE has demonstrated broad applicability in safeguarding data across various domains, including cloud computing, healthcare, and multi-party computation, further solidifying its role as a transformative technology for privacy-preserving operations~\cite{mollakuqe2024}. These studies demonstrate the broad potential of HE in safeguarding data throughout processing and communication. However, the application of HE for deep learning inference offloading on encrypted data in UAV–Edge systems remains an underexplored research direction.

\subsection{Challenges and Contributions}
\label{sec:I-C}
To address privacy limitations in UAV-based human face detection, we propose a novel framework that incorporates privacy-preserving techniques through Homomorphic Encryption (HE)~\cite{marcolla2022survey} to protect data privacy for the users and leverages mobile edge computing to mitigate intensive computing tasks for the UAV. In particular, in our proposed approach, when a drone/UAV captures a human face, it will immediately encrypt the image using the HE technique to protect human privacy. After that, the encrypted image will be sent to a nearby edge device for processing. On the edge side, an ML model is deployed to handle private image classification, performing encrypted human face detection without needing for decryption. 

Unlike conventional UAV–edge offloading schemes that typically transmit unencrypted sensor data or raw video streams with a focus on performance and energy efficiency, our framework redefines task offloading by treating all computational tasks as sensitive. Instead of merely offloading plain-text workloads for remote processing and transmitting raw information for centralized analysis, our approach encrypts each facial image immediately upon extraction on the device, ensuring that only encrypted text is transmitted to the edge server. The entire inference operation is encapsulated within a homomorphically encrypted domain, covering all stages from data collection, encryption on the UAV, and secure transmission to remote computation and returning encrypted results for decryption at the control station. By executing deep learning inference, including convolution operations, activation functions, and fully connected layers directly within the encrypted domain using the CKKS scheme, our secure offloading model not only leverages superior computational resources at the edge but also preserves UAV battery life, minimizes onboard latency, and maintains strict data security. The comprehensive integration of privacy-preserving mechanisms at every stage of the offloading process fundamentally distinguishes our work from prior studies that do not consider encryption as an intrinsic component of the offloading task.

    Nevertheless, despite the strength in
    cryptographic features (e.g., private computing, quantum-safe), these transformative advancements introduce several fundamental challenges. First, designing a model compatible with Homomorphic Encryption requires restructuring conventional two-dimensional convolutions into one-dimensional SIMD-based on multiplications, while replacing the standard activation functions such as ReLU or SiLU with Chebyshev polynomials of low degree. Second, effective noise budget management under the CKKS scheme demands meticulous arrangement of rotations, additions, and multiplications to ensure that accumulated noise remains within decryptable limits. Third, balancing security and performance necessitates evaluating different CKKS parameter configurations, employing ring sizes of 8192, 16384, and 32768 to achieve 256-bit security without incurring excessive computational costs or loss of accuracy. Homomorphic encryption inherently increases data size and computational overhead, posing additional optimization challenges. Furthermore, \textbf{HE only supports limited linear mathematical operations (element-wise multiplication and addition)}. Therefore, there is a critical need for approaches that can efficiently process homomorphically encrypted data at the edge.

Hence, it is challenging to directly apply HE functions to ML models, which requires various non-linear operations and matrix multiplication, especially for face detection problems.
To solve that problem, we first design a data encoding method in a Single-Instruction-Multiple-Data (SIMD) manner to convert raw 2D data into HE dimension, enabling effective encrypted convolution, which is challenging in HE-ciphertext. Building upon this encoding technique, we develop a private inference-based CNN algorithm on the encrypted image, allowing a privacy-prerserving human face detection approach. Moreover, observing \textbf{the challenge of implementing ML non-linear activation functions in HE}, we propose to leverage the Chebyshev approximation technique~\cite{lee2022cheb} to convert non-linear operations into polynomials with efficient degrees to human face data and HE dimensions. 
The main contributions of this work are summarized as follows:
\begin{itemize}
    \item To overcome privacy and resource constraints challenges in UAV-based human face detection, we propose a novel framework that integrates HE to ensure user data privacy while utilizing mobile edge computing to offload computationally intensive tasks from the UAV, enabling secure and efficient real-time processing.
    \item We develop an effective SIMD-based data encoding method for 2D convolution, allowing efficient convolution on HE-ciphertext. Following that, we design a CNN-based secure inference algorithm, effectively preserving users' privacy while performing face detection.
    \item We perform experiments on detection performance on a real-world drone-based human face detection dataset. The result indicates that our proposed method can achieve both highly accurate classification results and consistence performance compared to the non-encrypted baseline.
    \item We comprehensively evaluate the impact of polynomial approximation techniques on various activation functions under HE, including the use of Chebyshev-based functions of varying degrees. We also deeply analyze the effectiveness of different HE security levels, demonstrating the robustness and efficiency of our proposed algorithm for privacy-preserving human face detection across diverse post-quantum cryptographic settings.
\end{itemize}
% \textit{We need a few sentences to explain the proposed algorithm and how it can help to classify the encrypted image. Also, we need to highlight the key novel points of this approach as well as its benefits, e.g., high accuracy, fast speed, low complexity...}. Through intensive simulations on real-world dataset, we show that ... 

The rest of the paper is organized as follows. 
Section~\ref{sec:System-Model} presents our considered system model. Our proposed approach is then introduced in Section~\ref{sec:proposeddML}. Finally, the simulation results are discussed in Section~\ref{sec:Evaluation}, and Section~\ref{sec:Conclusion} concludes the work.

%===============================================
%===============================================
\section{System Model}
\label{sec:System-Model}

% \begin{figure*}[!]
%     \centering
%     \includegraphics[width=\linewidth]{Figs_remake/fig_proposed_system.pdf}
%     %\includegraphics[width=\columnwdith]{Figs_remake/fig_proposed_system_v2.pdf}
%     \caption{The proposed framework for face detection using UAV with privacy-preserving.}
%     \label{fig:proposed_framework}
%     %\vspace*{-0.5cm}
% \end{figure*}

Our proposed system model is presented in Fig.~\ref{fig:proposed_framework}. In particular, there are three main parties in this model, i.e., ($i$) a customer who wants to use a UAV-based face detection service (e.g., law enforcement and public safety agencies), ($ii$) a licensed UAV service provider who can deploy UAVs with high-quality cameras to serve requirements from the customer, and ($iii$) an edge AI provider who can deploy high-computing edge AI servers in a distributed manner to server users’ demands. In the beginning, the customer sends a service request to the UAV service provider and the Edge AI service provider. If they agree, the customer will share its \textbf{Homomophic Encryption (HE) public key} with them. This is to make sure that the service providers cannot access the results (i.e., people’s identification), which are sensitive and need to be protected. Specifically, when the UAV receives the customer's request, it will send a UAV equipped with a high-resolution camera for face-detection tasks to the serving area.

\begin{figure}[b]
    \centering
    \includegraphics[width=\columnwidth]{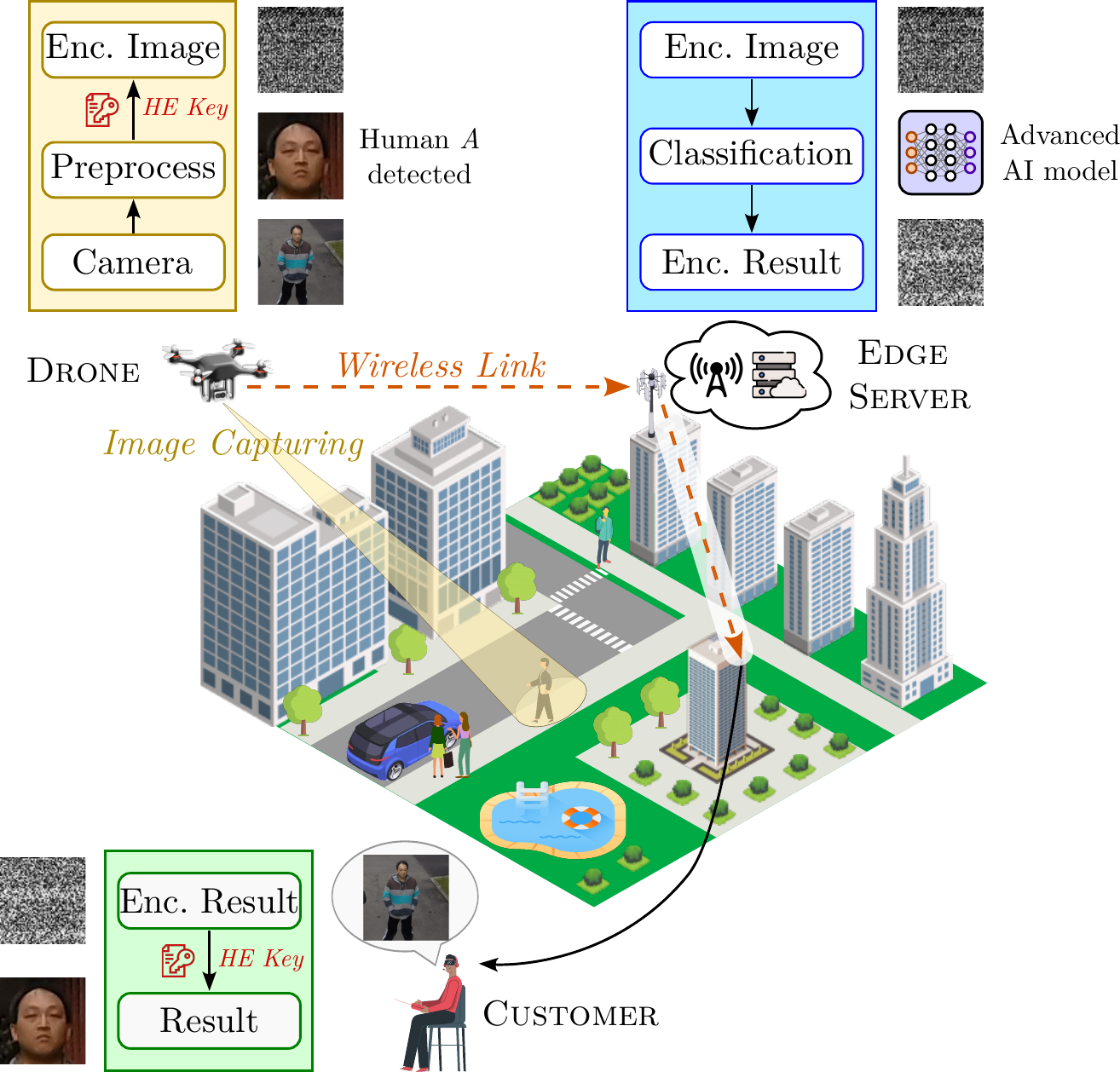}
    \caption{The Proposed Framework for Secure UAV-based Face Detection with HE and Edge AI.}
    \label{fig:proposed_framework}
    %\vspace*{-0.5cm}
\end{figure}

It is important to note that unlike conventional cameras (like mobile phone cameras), cameras equipped on drones for face detection require high resolution (e.g., 4K or higher) to capture fine facial details, wide dynamic range (WDR) to handle varying light conditions, and optical zoom for detecting faces from a distance~\cite{SAIWA_AI}. In addition, they require stabilization features, such as gimbal mounts, to ensure clear images despite drone motion. For example, the DJI Zenmuse H20T combines a 20 MP zoom camera with thermal imaging, making it ideal for long-range surveillance and facial recognition even in low light~\cite{Zenmuse_H20T}. Similarly, the Intel RealSense D435i adds depth sensing, enhancing accuracy in 3D face mapping for applications like crowd monitoring or secure access control~\cite{D435i}. These features are crucial for real-time, accurate, and reliable face detection in diverse operational scenarios.

When the UAV starts recording video from the serving area, instead of streaming all information to the edge server for processing (as this might consume significant energy and bandwidth due to extremely high-quality video; in addition, it would not be secure as sensitive information about people in the video could be exposed over wireless communication channels due to potential attacks such as eavesdropping, man-in-the-middle attacks, or jamming), the UAV will deploy lightweight on-device learning algorithms (e.g., YOLO~\cite{Chen2021YOLO} or MobileNet~\cite{Xu2021MobileNet}) to extract the faces of individuals in real-time. These detected faces will then be encrypted using an HE public key shared by the customer. In this case, the UAV only needs to transmit encrypted face images to the edge server for processing. This approach not only saves communication bandwidth and reduces latency but also protects user privacy and ensures a higher level of security by preventing direct exposure of sensitive data.

Beyond advancing privacy and encryption techniques, our work primarily examines task offloading by moving the computationally demanding deep learning inference from resource-constrained UAVs to powerful edge servers. Unlike previous approaches that stream raw sensor data or video frames for processing, we redefine the offloaded task as a complete homomorphically encrypted inference operation spanning data capture, onboard encryption, secure transmission, remote computation, and the return of encrypted results with sensitive inputs never being exposed. This treatment of encrypted inference as a first-class offloading unit leverages the superior computational resources available at the edge while simultaneously preserving UAV battery life, minimizing onboard latency, and ensuring stringent data confidentiality. Ultimately, our approach addresses the inherent limitations of UAV platforms and extends the classical offloading paradigm into a secure domain that preserves privacy for edge computing.

% Beyond advancing privacy and encryption techniques, our work primarily reexamines task offloading by moving the computationally demanding deep learning inference from resource-constrained UAVs to powerful edge servers. Unlike previous approaches that stream raw sensor data or video frames for processing, we redefine the offloaded task as a complete homomorphically encrypted inference operation spanning data capture, onboard encryption, secure transmission, remote computation, and the return of encrypted results with sensitive inputs never being exposed. This treatment of encrypted inference as a first class offloading unit leverages the superior computational resources available at the edge while simultaneously preserving UAV battery life, minimizing onboard latency, and ensuring stringent data confidentiality. Ultimately, our approach addresses the inherent limitations of UAV platforms and extends the classical offloading paradigm into a secure domain that preserves privacy for edge computing.}

Once the Edge AI Server (EAIS) receives the encrypted image from the UAV, the proposed advanced AI model will be executed to perform the classification task. The EAIS in the system serves as a critical intermediary that offloads computationally intensive tasks from the UAV while ensuring data privacy and security. Operating in an untrusted environment, the edge server receives encrypted images from the UAV, which are processed without decryption using our proposed approach in Section~\ref{sec:proposeddML} to maintain privacy. Using our proposed AI model, which is compatible with HE, the EAIS can perform face classification tasks directly on the encrypted data. Here, it is noted that the result from the classification task is encrypted, which means that the EAIS does not know the exact results (i.e., people’s identities). This thus ensures that sensitive information remains protected, even if the server is compromised. After processing, the EAIS generates encrypted results, which are transmitted back to the customer or UAV for decryption and further use. By processing data closer to the source, the edge server reduces the bandwidth needed for transmitting raw video, minimizes latency, and enables real-time applications, all while adhering to strong privacy guarantees. 

Finally, when the encrypted results are sent to the customer, they will decrypt it using the customer’s HE private key to obtain the final results (i.e., people’s identities). It is important to note that through the whole process, the people’s identities remain secured, and only the customer (e.g., law enforcement and public safety agencies) can access the people’s identities. All other providers (e.g., UAV and EAIS providers) cannot access any information about the people’s identities (although they can access the raw data, e.g., the UAV provider can access the human face, but they have no idea who they are).

%=================================================
%=================================================
\section{Proposed ML Model for Privacy-Preserving Inference with Homomorphic Encryption}
\label{sec:proposeddML}

\subsection{Homomorphic Encryption}
\label{sec:3a}

%\begin{figure}[t]
%    \centering
%    \includegraphics[width=\linewidth]{figures/enc_weight.pdf}
%    \caption{Encrypt process of a 3x3 matrix \textbf{\textit{W}}.}
%    \label{fig:encW}
%\end{figure}

\begin{table}[t]
    \centering
    \footnotesize
    \renewcommand{\arraystretch}{1.3}
    \caption{Summary of Notations.}
    \label{tab:notations}
    % \resizebox{0.5\textwidth}{!}{     
     \begin{tabular}{cl}
        \toprule 
        \textbf{Symbol} & \textbf{Description} \\
        \cmidrule[0.4pt](lr{0.12em}){1-1}%
        \cmidrule[0.4pt](lr{0.12em}){2-2}%
        $B$ & Ciphertext size in CKKS \\
        $R$ & Ring dimension in CKKS \\
        $\mathsf{sk}_n$ & Secret key for user $n$ \\
        $\mathsf{pk}_n$ & Public key for user $n$ \\
        $c$ & Plaintext before encryption \\
        $\overline{c}$ & Ciphertext after encryption \\
        $\overline{c}_{\text{add}}$ & Ciphertext result after addition \\
        $\overline{c}_{\text{mult}}$ & Ciphertext result after multiplication \\
        $H, W$: & Height and width of the input matrix \\
        $k_H, k_W$ & Height and width of the convolution kernel \\
        $s_H, s_W$ & Stride parameters (height, width) \\
        $p_H, p_W$ & Padding parameters (height, width) \\
        $H_o, W_o$ & Output height and width of convolution \\
        $j_H$, $j_W$ & Row and column indices of the \\ 
        & sliding convolution window \\
        $h_k^{(i)}, w_k^{(i)}$ & Relative row and column positions \\
        & of a kernel element \\
        $I_{(encode)}$ & Encoded input matrix for encrypted convolution \\
        $I_{\text{plain}}$ & Flattened and zero-padded plaintext for CKKS \\
        $K_{\text{plain}}$ & Encoded kernel as CKKS plaintext \\
        $\overline{I}$ & Encrypted input matrix in CKKS \\
        $\overline{O}_{\text{conv}}$ & Encrypted convolution output \\
        $F C$ & Fully connected layer matrix \\
        $FC_{\text{plain}}^{(k)}$  & Encoded fully connected layer as CKKS plaintext \\
        $\overline{x}^{(k)}$ & Encrypted input to the $k$-th linear layer \\
        $\overline{\otimes}$ & Encrypted element-wise multiplication \\
        $\mathsf{Enc}(\mathsf{pk}_n, c)$ & Encryption function \\
        $\mathsf{Dec}(\mathsf{sk}_n, \overline{c})$ & Decryption function \\
        $\mathsf{Add}(\overline{c}_1, \overline{c}_2)$ & Ciphertext addition \\
        $\mathsf{PlAdd}(\overline{c}_1, c_2)$ & Plaintext-ciphertext addition \\
        $\mathsf{Mult}(\overline{c}_1, \overline{c}_2)$ & Ciphertext multiplication \\
        $\mathsf{PlMult}(\overline{c}_1, c_2)$ & Plaintext-ciphertext multiplication \\
        $\mathsf{Rot}(\overline{c}, i)$ & Rotation operation on ciphertext \\
        $\mathsf{ZeroPad}$ & Zero padding function \\
        $\mathsf{Flatten}$ & Flattening function \\
        $\mathsf{Replicate}$ & Replication function for kernel encoding \\
        $\mathsf{DiagonalEncode}$ & Encoding function for fully connected layers \\
        \bottomrule
     \end{tabular}
    % } % Đóng resizebox trước caption
\end{table}

%Manh's part
Homomorphic encryption (HE) is an advanced cryptography approach that allows computations directly on encrypted data without needing for decryption~\cite{gentry2009fully}. In this work, we leverage the Cheon-Kim-Kim-Song (CKKS) scheme, which supports encrypted calculation on real (floating-point) numbers, making it suitable to integrate with deep learning models~\cite{cheon2017ckks}. Particularly, CKKS requires the ciphertext to be a one-dimensional (1D) vector with a size of $B = R/2$, where 
$B$ is the ciphertext size and $R$ is the ring dimension used to ensure the HE security level and multiplicative depth~\cite{gentry2009fully},~\cite{cheon2017ckks}. In CKKS, the raw data need to be encoded into 1D plaintext before encrypting into the ciphertext. To be more specific, the CKKS provides basic HE evaluation algorithms as follows:
\begin{itemize}
    \item \textbf{Key generation}: Generate random secret key $\mathsf{sk}_n$ and public key $\mathsf{pk}_n$ for user $n$.
    % \item \mathsf{SKGen}$(n)$: generate random secret key $\mathsf{sk}_n$ for user $n$.
    % \item $\mathsf{PKGen}(\mathsf{sk}_n)$: create the public key $\mathsf{pk}_n$ for user $n$ based on the secret key $\mathsf{sk}_n$.
    \item \textbf{Encryption}: Given a 1D plaintext $c$ and $\mathsf{pk}_n$, encrypt $c$ into a ciphertext $\overline{c}$ which $\overline{c} = \mathsf{Enc}(\mathsf{pk}_n, c)$.
    \item \textbf{Decryption}: Given a ciphertext $\overline{c}$ and $\mathsf{sk}_n$, decrypt $\overline{c}$ into its plain form $c$ which $c = \mathsf{Dec}(\mathsf{sk}_n, \overline{c})$.
    \item \textbf{Addition}: Given two ciphertext $\overline{c}_1$ and $\overline{c}_2$, perform element-wise encrypted addition $\mathsf{Add}(\overline{c}_1,\overline{c}_2) = \overline{c}_{\text{add}}$, in which $\mathsf{Dec}(\mathsf{sk}_n, \overline{c}_{\text{add}}) \approx c_1 + c_2$.
    \item \textbf{Addition by Plaintext}: Given a ciphertext $\overline{c}_1$ and a plaintext $c_2$, perform element-wise encrypted addition $\mathsf{PlAdd}(\overline{c}_1, c_2) = \overline{c}_{\text{add}}$, in which $\mathsf{Dec}(\mathsf{sk}_n, \overline{c}_{\text{add}}) \approx c_1 + c_2$.
    \item \textbf{Multiplication}: Given two ciphertext $\overline{c}_1$ and $\overline{c}_2$, perform element-wise encrypted multiplication $\mathsf{Mult}(\overline{c}_1,\overline{c}_2) = \overline{c}_{\text{mult}}$, in which $\mathsf{Dec}(\mathsf{sk}_n, \overline{c}_{\text{mult}}) \approx c_1 \times c_2$.
    \item \textbf{Multiplication by Plaintext}: Given a ciphertext $\overline{c}_1$ and a plaintext $c_2$, perform element-wise encrypted multiplication $\mathsf{PlMult}(\overline{c}_1, c_2) = \overline{c}_{\text{mult}}$, in which $\mathsf{Dec}(\mathsf{sk}_n, \overline{c}_{\text{mult}}) \approx c_1 \times c_2$.
    \item \textbf{Rotation}: Given a ciphertext $\overline{c}$ encrypting a raw vector $(c_1, \dots, c_{(N/2)-1})$, perform $i$-th rotation $\mathsf{Rot}(\overline{c}, i)$ = $\overline{c}_{Rot}$, in which $\mathsf{Dec}(\mathsf{sk}_n, \overline{c}_{Rot}) \approx (c_i,\dots, c_{(N/2)-1}, c_1\dots, c_{i-1})$. In this work, we consider the $\mathsf{Rot}(.)$ as left-rotation.
    % \item $\mathsf{Mult}(\bm{\hat{c}_1},\bm{\hat{c}_2})$: the multiplication between two ciphertexts $\bm{\hat{c}_1}$ and $\bm{\hat{c}_2}$, in which $\mathsf{Dec}(\mathsf{sk}_n, \mathsf{Mult}(\bm{\hat{c}_1},\bm{\hat{c}_2})) \approx \mathbf{c_1} \times \mathbf{c_2}$.
\end{itemize}

% The procedure of the HE algorithm is summarized in \lqt{Fig.~\ref{X}} \lqt{Trung: Make a block diagram for HE below}.
It is worth noting that in CKKS, addition and multiplication can be performed element-wise between ciphertext-ciphertext or plaintext-ciphertext. In this work, we define these operations separately to provide a clearer understanding of the formulation. The symbols used in the paper are presented in Table~\ref{tab:notations}.

\subsection{Proposed DL Model for HE-encrypted Data}
\label{sec:3b}

\begin{figure*}[!t]
    \centering
    \includegraphics[width=\textwidth]{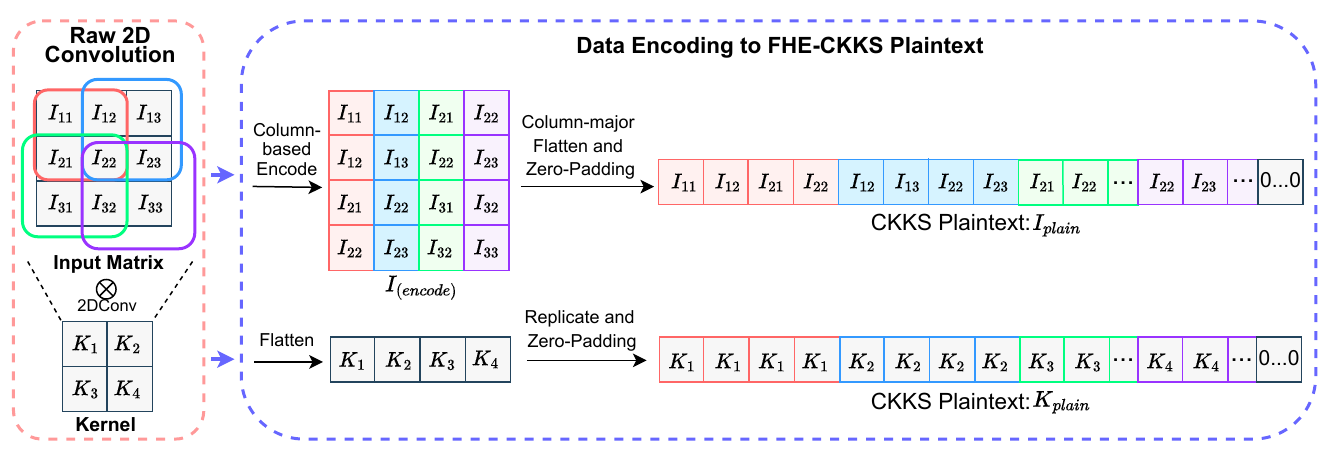}
    \caption{The proposed data encoding technique. The input matrix is first reorganized into a column-based matrix $I_{(encode)}$. Then, it is flattened and zero-padded into 1D CKKS format. The kernel is also flattened and replicated to match the CKKS input data for encrypted convolution.}
    \label{fig:im2col}
\end{figure*}

% \lqt{Trung: Better providing a summary of all steps of the proposed approach.}
\subsubsection{The SIMD-based Data Encoding}
As mentioned in Section~\ref{sec:System-Model}, the DL model deployed on the EAIS is required to perform face-detection tasks on encrypted images transmitted from the UAV. Therefore, we leverage a convolutional neural network (CNN) specifically designed for encrypted data inference to effectively classify the incoming encrypted images. In particular, the proposed CNN contains a 2D convolutional layer and a fully connected layer. Observing the challenges of performing convolution on encrypted data, such as the limitations of element-wise operations and the 1D format of CKKS ciphertexts, we develop an encoding technique designed to preprocess raw data (e.g., input matrix and CNN kernels) into CKKS plaintext/ciphertext, ensuring efficient encrypted convolution after CKKS encryption. By following the Single Instruction Multiple Data (SIMD) manner, the encoding technique allows the 2D convolution of CNN to operate as a single encrypted matrix multiplication, making it particularly compatible with CKKS. Generally, Fig.~\ref{fig:im2col} describes the processes of our SIMD-encoding method. Given the input matrix $I \in \mathbb{R}^{H\times W}$ and kernel $K \in \mathbb{R}^{{k_H}\times {k_W}}$ with the stride parameters ($s_H, s_W$) and padding parameters ($p_H, p_W$). Our encoding technique initially generates the output matrix $I_{(encode)} \in  \mathbb{R}^{({k_H}\cdot {k_W})\times (H_o\cdot W_o)}$, where each column represents a convolutional window extracted from the input matrix. Here, $H_o$ and $W_o$ are the height and width of the output of the unencrypted 2D convolution, which can be defined as:
\begin{equation}
    H_o = \frac{H + 2p_h -k_H}{s_h} + 1;
    W_o = \frac{W + 2p_w -k_W}{s_w} + 1.
\end{equation}

% \begin{figure*}[!t]
%     \centering
%     % \includegraphics[width=\linewidth]{Figs/convolution_im2col b2.png}
%     \includegraphics[width=.9\textwidth]{Figs_remake/fig_im2col.pdf}
%     \caption{im2col \lqt{Trung: Add full caption.}.}
%     \label{fig:ckks_transform}
% \end{figure*}

Accordingly, the output matrix $I_{(encode)}$ is calculated based on the column representation of the convolution windows, which can be defined as:
\begin{equation}
\label{eq:im2col}
\begin{aligned}
    I_{(encode) i,j} &= \mathsf{2DCol}(I), \\
    &= I_{(j_H\cdot s_H +h_k^{(i)}-p_H), (j_W\cdot s_W+w_k^{(i)}-p_W)},
\end{aligned}
\end{equation}
where $j_H$ and $j_W$ denote the respective row and column indices of the sliding convolution window on the input matrix, representing the global position of the convolution window from the top-left corner of input. Particularly, they can be calculated as:
\begin{align}
    j_H &= \left\lfloor \frac{j}{W_o} \right\rfloor, &&\text{where}~0 \leq j_H < H_o,\\
    j_W &= j \bmod W_o, &&\text{where}~0 \leq j_W < W_o.
\end{align}
Meanwhile, in \eqref{eq:im2col}, $h_k^{(i)}$ and $w_k^{(i)}$ represent the relative row and column positions of the kernel element within the sliding convolution window on the input matrix, which can be defined as:
\begin{equation}
    h_k^{(i)} = \left\lfloor \frac{i}{k_W} \right\rfloor, w_k^{(i)} = i \bmod k_W,
\end{equation}
in which $ h_k^{(i)}$ and $w_k^{(i)}$ are fixed for the kernel and determine how each kernel element aligns with the input matrix during convolution. By combining ($j_H$, $j_W$) and ($h_k^{(i)}$, $w_k^{(i)}$), we can reorganize the input matrix into a column-based, mapping each column of the encoded matrix with the corresponding kernel elements for accurate SIMD encrypted convolution. To be more specific, Fig.~\ref{fig:im2col} illustrates the general transformation from the input matrix into the column-based form using~(\ref{eq:im2col}). Subsequently, the encoded matrix $I_{(encode)}$ is converted into CKKS plain form:
\begin{equation}
    I_{\text{plain}} = \mathsf{ZeroPad}(\mathsf{Flatten}(I_{(encode)}),B),
\end{equation}
where $B$ is the ciphertext size, as mentioned in section~\ref{sec:3a}. The $I_{(encode)}$ is first flattened into the 1D format and then zero-padded to match the size of the CKKS ciphertext, as illustrated in Fig.~\ref{fig:im2col}. Then, using the public key $\mathsf{pk}$ generated from the user side, the plaintext $I_{\text{plain}}$ can be encrypted into CKKS ciphertext, which can be defined as:
\begin{equation}
    \overline{I} = \mathsf{Enc}(I_{\text{plain}}, \mathsf{pk}).
\end{equation}

Regarding the kernel of CNN, it is encoded into CKKS plaintext which is compatible with the input ciphertext $\overline{I}$ during convolution. The encoding process of the kernel is defined as:
\begin{equation}
    K_{\text{plain}} = \mathsf{Replicate}(\mathsf{Flatten}(K), (H_o\times W_o)),
\end{equation}
where $(H_o\times W_o)$ denotes the number of kernel windows needed to produce the output based on raw convolution. Initially, the kernel is flattened into the 1D format to match with the input CKKS ciphertext. 
% Then, its element is replicated $(H_o\times W_o)$ times to ensure efficient encrypted convolution with the input ciphertext, as illustrated in Fig.~\ref{fig:im2col}. 
Then, its element is replicated, and the kernel elements are systematically arranged into clusters, as illustrated in Fig.~\ref{fig:im2col}, to align with the positions of corresponding elements in the input ciphertext \( I_{\text{plain}} \). Specifically, after flattening, each kernel element is duplicated consecutively, precisely matching the number of convolution windows (\( H_o \times W_o \)). This ensures that during the multiplication between the image ciphertext and the plaintext kernel, each kernel element correctly aligns with the corresponding image elements in each window. This arrangement is a necessary condition for accurately and efficiently performing encrypted convolution, as the positions of ciphertext elements cannot be altered after encryption. Structuring the kernel in this manner allows us to leverage the wide-parallel computation capabilities (SIMD) of the CKKS scheme, optimizing encrypted convolution operations.
% Then, its element is replicated, and the kernel elements are systematically arranged into clusters, as illustrated in Fig.~\ref{fig:im2col}, to align with the positions of corresponding elements in the input ciphertext \( I_{\text{plain}} \). Specifically, after flattening, each kernel element is duplicated consecutively, precisely matching the number of convolution windows (\( H_o \times W_o \)). This ensures that during the multiplication between the image ciphertext and the plaintext kernel, each kernel element correctly aligns with the corresponding image elements in each window. This arrangement is a necessary condition for accurately and efficiently performing encrypted convolution, as the positions of ciphertext elements cannot be altered after encryption. Structuring the kernel in this manner allows us to leverage the wide-parallel computation capabilities (SIMD) of the CKKS scheme, optimizing encrypted convolution operations.

\begin{figure*}[!t]
    \centering
    \includegraphics[width=\linewidth]{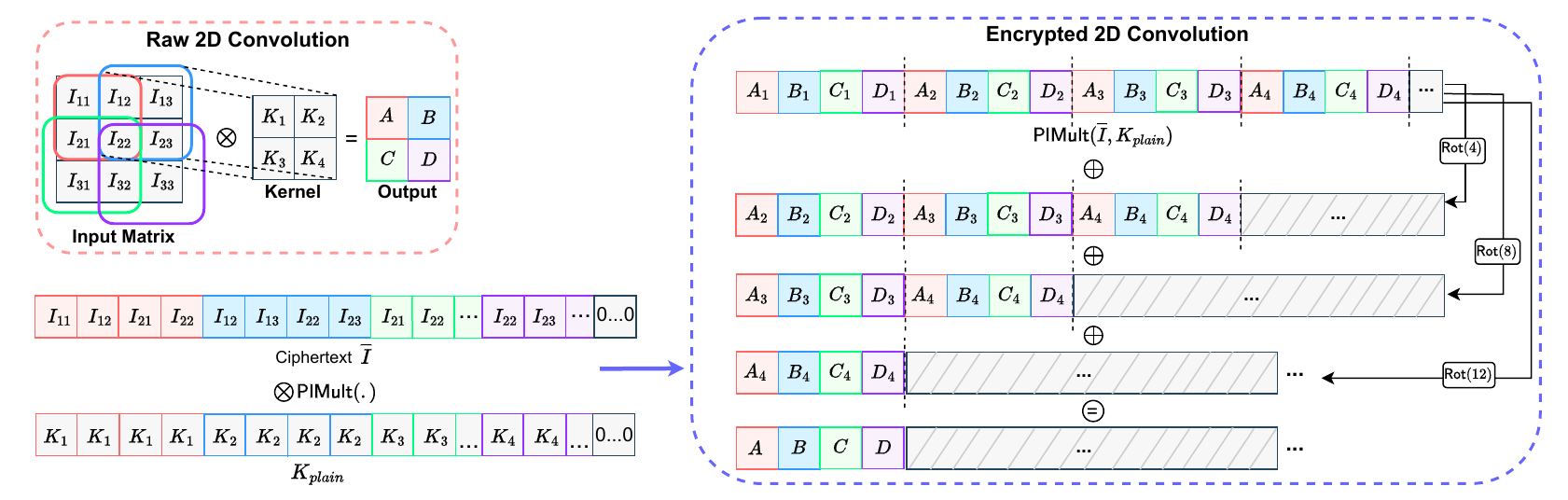}
    \caption{Illustration of the designed encrypted convolution. The input ciphertext is first multiplied with encoded plaintext-kernel. The output is then HE-rotated and summed to obtain the final convolution result.}
    \label{fig:conv2d}
    \vspace*{-0.2cm}
\end{figure*}

\subsubsection{The Secure CKKS-based CNN Inference}
After finishing the encoding/encryption of the input matrix and kernel of the CNN, the output $\overline{O}$ of encrypted convolution can be calculated as:
\begin{equation}
\label{eq:enc_conv}
    \overline{O}_{\text{conv}} = \sum_{r=0}^{k_H\cdot k_W - 1} \mathsf{Rot}\Big(\mathsf{PlMult}(\overline{I}, K_{\text{plain}}),(r\cdot (H_o\cdot W_o))\Big),
\end{equation}
where the input ciphertext $\overline{I}$ and plain kernel $K_{\text{plain}}$ are first performed element-wise multiplication by plaintext. The output ciphertext, after performing $\mathsf{PlMult}(.)$, is then left-rotated into multiple versions, which are subsequently summed into the final output ciphertext $\overline{O}_{\text{conv}}$, as illustrated in Fig.~\ref{fig:conv2d}.

Regarding the fully connected layer (FCL) of the CNN, we leverage the diagonal encoding from Halevi and Shoup~\cite{diagonal} to transform the linear layers into CKKS plaintexts, which is defined as:
\begin{equation}
    FC_{\text{plain}} = \mathsf{DiagonalEncode}(FC),
\end{equation}
where $FC$ and $FC_{\text{plain}}$ are the linear layer and its correspond CKKS plaintext, respectively. Hence, the encrypted computation through the linear layers is defined as:
\begin{equation}
    \overline{x}^{(k+1)} = \overline{x}^{(k)}~\overline{\otimes}~FC_{\text{plain}}^{(k)}.
\end{equation}
where $FC_{\text{plain}}^{(k)}$ and $\overline{x}^{(k)}$ are the CKKS plaintext and input ciphertext of the $k$-th linear layer, respectively. The $\overline{\otimes}$ is adopted from~\cite{diagonal} which perform multiple $\mathsf{PlMult}(.)$ in diagonal manner. For clarity, the diagonal algorithm is generally demonstrated in Fig.~\ref{fig:linear}.
It is worth noting that in this work, the input to the first linear layer $\overline{x}^{(1)}$ is equal to $\overline{O}_{\text{conv}}$ due to our CNN configuration, which is described in Section~\ref{sec:4a}. 

\begin{algorithm}[htb]
\caption{Privacy-Preserving Inference on Encrypted Facial Data}
\KwIn{Face image $I \in \mathbb{R}^{H \times W}$, with weights $\{W_{\text{conv}}, \text{FC}\}$}
\KwOut{Encrypted classification result $\hat{y}$}

Resize and normalize $I$;

% \textit{\# Convert to column using (\ref{eq:im2col})}

$I_{\text{encoded}} \leftarrow \mathsf{2DCol}(I)$;   \Comment{Convert to column using (\ref{eq:im2col})}

$I_{\text{plain}} \leftarrow \mathsf{ZeroPad}(\mathsf{Flatten}(I_{\text{encoded}}), B)$;

$\bar{I} \leftarrow \mathsf{Enc}(I_{\text{plain}}, pk)$; \Comment{Encrypt with CKKS}

$K_{\text{plain}} \leftarrow \mathsf{Replicate}(\mathsf{Flatten}(W_{\text{conv}}), H_o \cdot W_o)$;

\textit{\# SIMD-based convolution using (\ref{eq:enc_conv})}

$\bar{O}_{\text{conv}} \leftarrow \mathsf{EncryptedConv}(\bar{I}, K_{\text{plain}})$;  %\Comment{SIMD-based convolution using (\ref{eq:enc_conv})}\;

\ForEach{FC$^{(k)}$ in FC layers}{
    $FC^{(k)}_{\text{plain}} \leftarrow \mathsf{DiagonalEncode}(FC^{(k)})$;

    $\bar{x}^{(k+1)} \leftarrow \bar{x}^{(k)} \otimes FC^{(k)}_{\text{plain}}$; \Comment{Encrypted linear}
}

\Return $\hat{y} \leftarrow \bar{x}^{(\text{final})}$;

\end{algorithm}

% \begin{figure*}[!t]
%     \centering
%     \begin{subfigure}[b]{0.7\linewidth}
%         \centering
%         \includegraphics[width=\linewidth]{Figs_remake/fig_linear_step_1.pdf}
%         \caption{Draft encrypted linear \lqt{Trung: Add full caption for step 1}.}
%         \label{fig:linear_step1}
%     \end{subfigure}
%     \hfill
%     \begin{subfigure}[b]{0.8\linewidth}
%         \centering
%         \includegraphics[width=\linewidth]{Figs_remake/fig_linear_step_2.pdf}
%         \caption{Draft encrypted linear \lqt{Trung: Add full caption for step 2}.}
%         \label{fig:linear_step2}
%     \end{subfigure}
%     \caption{Encrypted linear process illustrated in two steps. \lqt{Trung: Update with proper description.}}
%     \label{fig:linear_combined}
%     \vspace*{-0.2cm}
% \end{figure*}

% \begin{figure*}[!t]
%     \centering
%     % \includegraphics[width=\linewidth]{Figs/linear b1.png}
%     \includegraphics[width=.9\linewidth]{Figs_remake/fig_linear_step_1.pdf}
%     \caption{Draft encrypted linear \lqt{Trung: Add full caption}.}
%     \label{fig:linear}
%     \vspace*{-0.2cm}
% \end{figure*}

% hình gốc
% \begin{figure*}[!t]
%     \centering
%     % \includegraphics[width=\linewidth]{Figs/linearb2.png}
%     \includegraphics[width=\linewidth]{Figs_remake/fig_linear_step_2.pdf}
%     \caption{Illustration of the diagonal algorithm adopted from~\cite{diagonal}.}
%     \label{fig:linear}
%     \vspace*{-0.2cm}
% \end{figure*}

\begin{figure*}[!t]
    \centering
    \includegraphics[width=\linewidth]{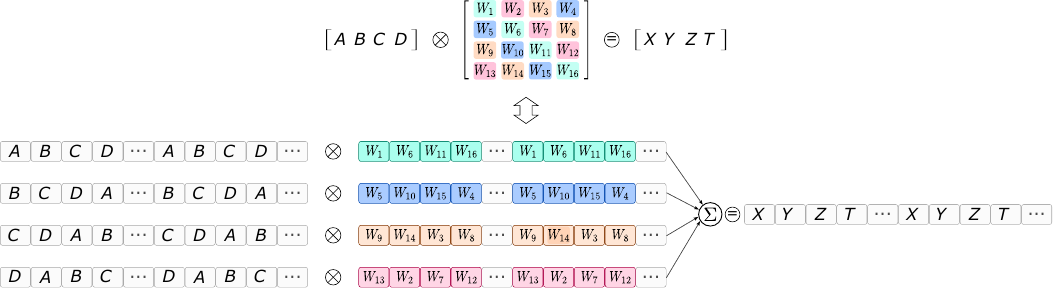}
    \caption{Illustration of the diagonal algorithm adopted from~\cite{diagonal}.}
    \label{fig:linear}
    \vspace*{-0.2cm}
\end{figure*}

\section{Performance Evaluation}
\label{sec:Evaluation}
%-----------------------------------
%-----------------------------------
\subsection{Simulation Setup}
\label{sec:4a}
In this section, we conduct comprehensive simulations to evaluate the performance of the proposed framework. The proposed algorithm is implemented using the CKKS scheme of the TenSEAL library~\cite{tenseal}, configured with a ring dimension of $8192$, $16384$, and $32768$ and the corresponding HE batch sizes of $4096$, $8192$, and $16384$ are selected to ensure $256$-bit security, protecting against quantum attacks~\cite{guo2024key}. Increasing security levels requires a larger ring dimension, which directly leads to an increase in batch size due to the algorithmic constraints of the HE encoding scheme. This design choice aligns with our encoding approach and allows us to thoroughly evaluate detection accuracy at different encryption levels, ensuring the robustness of our system under various security settings. To validate the effectiveness of the proposed privacy protection model, we leverage the DroneFace dataset~\cite{hsu2017droneface}, a well-established open dataset specifically designed to assess the performance of facial recognition algorithms in UAV-based applications and their applicability to real-world UAV surveillance scenarios. The detailed descriptions of the dataset, along with its pre-processing steps, are provided in the next section.

% Regarding the proposed deep neural network (DNN), we designed a fully connected network comprising an input layer, two hidden layers, and an output layer. The respective layers contained $784$, $256$, $64$, and $11$ neurons. In addition to the input layer, each layer was appended with the ReLU activation function.

In the proposed deep neural network (DNN), we designed a convolutional neural network (CNN) consisting of one convolutional layer followed by two fully connected layers. The model includes a single convolutional layer with 4 output channels, using a $7 \times 7$ kernel, a stride of $3$, and no padding. The output of the convolutional layer is flattened and passed through a fully connected layer with $64$ neurons, followed by another fully connected layer with $11$ output neurons for classification. In addition to the input layer, each layer was appended with the activation function. The model employs the cross-entropy loss function $\mathsf{CrossEnrtropyLoss(.)}$ to evaluate the performance of the classification task. For optimization, the Adam optimizer is utilized with a learning rate of $0.001$.

% \lqt{we need more details here ... how about batch size? learning rate ...}. 
%Beyond the input layer, each layer incorporates an activation function such as ReLU, SiLU, or Square. The model employs the cross-entropy loss function $$\mathsf{CrossEnrtropyLoss(.)}$$ to evaluate the performance of the classification task. For optimization, the Adam optimizer is utilized with a learning rate of $0.001$. A batch size of $32$ and $1$ is used for the training and testing, respectively.

% During training, a batch size of $32$ is used, while a batch size of $1$ is adopted for testing.

To evaluate the efficiency of the proposed solution, we compare its performance against a benchmark ML algorithm operating on \emph{non-encrypted} data, which serves as the theoretical upper bound for accuracy. This aligns with prior works on secure neural network inference and privacy-preserving AI techniques, such as encrypted neural network inference for human action recognition~\cite{kim2022secure}, privacy-preserving cyberattack detection in blockchain-based IoT systems~\cite{manh2025privacy}, and HE-encrypted training frameworks based on federated learning~\cite{nguyen2024encrypted},~\cite{hieu2024vtc}. The proposed approach is expected to achieve a close classification accuracy for facial images to that yielded by this benchmark, demonstrating its practical viability. To ensure a fair and consistent comparison, both the benchmark and the proposed approach utilize the same CNN architecture, eliminating potential biases stemming from architectural differences and providing a clear evaluation of the impact of encryption on model performance. As mentioned in Section~\ref{sec:I-C}, since adopting non-linear activation functions in HE is challenging, we evaluate our proposed CNN-based algorithm with different activation functions, including advanced ReLU, SiLU, and a lightweight Square function. Note that these activation functions are approximated using the Chebyshev polynomial approach, in which its assessment is shown in Section~\ref{sec:D-II}.
 %In addition, we provide comparisons with a CNN using the square function. This benchmark will better show the efficiency of using the ReLU function in our proposed approach.   

%-----------------------------------
%-----------------------------------
\subsection{Dataset and Preprocessing}
%\cite{DroneFaceRoboflow} 
We utilize the DroneFace dataset from~\cite{hsu2017droneface} for the performance evaluation. This dataset includes $11$ classes, consisting of seven males and four females, all Taiwanese, $23$ to $36$ years old, with heights ranging from $157$ to $183$ cm. Each subject was photographed in frontal and tilted portrait poses before the raw image collection, resulting in $1,437$ images, including $73$ portrait images, $747$ frontal images, and $620$ raw images of the $11$ subjects. The resolution of facial images ranges from $23 \times 31$ to $384 \times 384$. Three-dimensional portrait images were captured using both sports cameras and phone cameras for comparison. Raw images were taken at heights of $1.5$, $3$, $4$, and $5$ meters, at distances from $2$ to $17$ meters from the subjects, with $0.5$-meter intervals. The $11$ subjects are labeled with English letters from $a$ to $k$, with subjects $a$, $b$, $c$, $e$, $g$, $j$, and $k$, being male and the remaining subjects being female. The dataset was then divided into training and testing sets with a ratio of $80:20$. The preprocessing steps included resizing to $28 \times 28$ pixels and normalizing to the range $[-1,1]$. The images were converted to grayscale with a single channel, transformed into PyTorch tensors, and normalized with a mean of $0.5$ and a standard deviation of $0.5$.

%-----------------------------------
%-----------------------------------
\subsection{Performance Metric}

To evaluate the performance of the detection model, four  macro-average standard metrics for classification problems are used: ($i$) the accuracy which is the percentage of correctly classified (true positives and true negatives) instances, ($ii$) the precision which measures the proportion of true positive predictions among all instances classified as positive, ($iii$) the recall which quantifies the percentage of true positive instances among all actual positive instances, and ($iv$) the confusion matrix which is used to evaluate the performance of a classification model by comparing predicted and actual values. Expression of the first three metrics are given below:
\begin{subequations}
    \begin{align}
        \text{Accuracy} &= 
            \frac{1}{C} \sum_{c=1}^{C} \frac{\text{TP}_c + \text{TN}_c}{\text{TP}_c + \text{TN}_c + \text{FP}_c + \text{FN}_c}, \\
        \text{Precision} &= 
            \frac{1}{K} \sum_{k=1}^K \frac{\text{TP}_k}{\text{TP}_k + \text{FP}_k}, \\
        \text{Recall} &= 
            \frac{1}{K} \sum_{k=1}^K \frac{\text{TP}_k}{\text{TP}_k + \text{FN}_k },
    \end{align}
\end{subequations}
where $K$ is the number of classes in the system and TP, TN, FP, and FN stand for true positive, true negative, false positive, and false negative, respectively.

% the confusion matrix is utilized, which is suitable for a machine learning-based classification system. 
% \begin{equation} 
%     \text{Accuracy} = \frac{1}{C} \sum_{c=1}^{C} \frac{\text{TP}_c + \text{TN}_c}{\text{TP}_c + \text{TN}_c + \text{FP}_c + \text{FN}_c}.
% \end{equation}

% The macro-average precision and recall are utilized in this term. Given $K$ as the number of classes in the system, the macro-average precision is:
% \begin{equation}
%     \text{Precision} = \frac{1}{K} \sum_{k=1}^K \frac{\text{TP}_k}{\text{TP}_k + \text{FP}_k}.
% \end{equation}

% The macro-average recall is calculated as follows:
% \begin{equation}
%     \text{Recall} = \frac{1}{K} \sum_{k=1}^K \frac{\text{TP}_k}{\text{TP}_k + \text{FN}_k }.
% \end{equation}

% A confusion matrix is a table used to evaluate the performance of a classification model by comparing the predicted values against the actual values.

%\usepackage{float} % Thêm gói này vào phần đầu tài liệu

%-----------------------------------
%-----------------------------------
\subsection{Simulation Results}

\subsubsection{Convergence Analysis}

%Đức viết
Fig.~\ref{fig:accuracy} presents the accuracy of our proposed model when using three activation functions: ($i$) ReLU and ($ii$) SiLU, and ($iii$) Square for both non-encrypted and encrypted data. As observed, the models employing ReLU and SiLU functions reach stability and comparable accuracy for both non-encrypted and encrypted data after approximately $20$ training rounds. In contrast, models using the square function exhibit stability after around $30$ epochs. Despite slight variations in convergence speed, the learning accuracy of the model using all three functions remains analogous, stabilizing between $93\%$--$95\%$.

These results highlight that the square function initially demonstrates lower accuracy, which improves after $20$ epochs. However, it continues to exhibit fluctuations and does not attain the high accuracy levels achieved by ReLU and SiLU. Subsequently, the SiLU function, initially showing lower accuracy compared to ReLU, gradually catches up, achieving nearly equivalent accuracy. The SiLU activation function is characterized by its smoother transition in comparison to ReLU. The ReLU function demonstrates robust performance and rapid convergence, making it an optimal choice for deep neural networks.

%Proposed method using relu function.
\begin{figure}[!]
    \centering
    % First subfigure
    \begin{subfigure}{0.35\textwidth}
        \includegraphics[width=\textwidth]{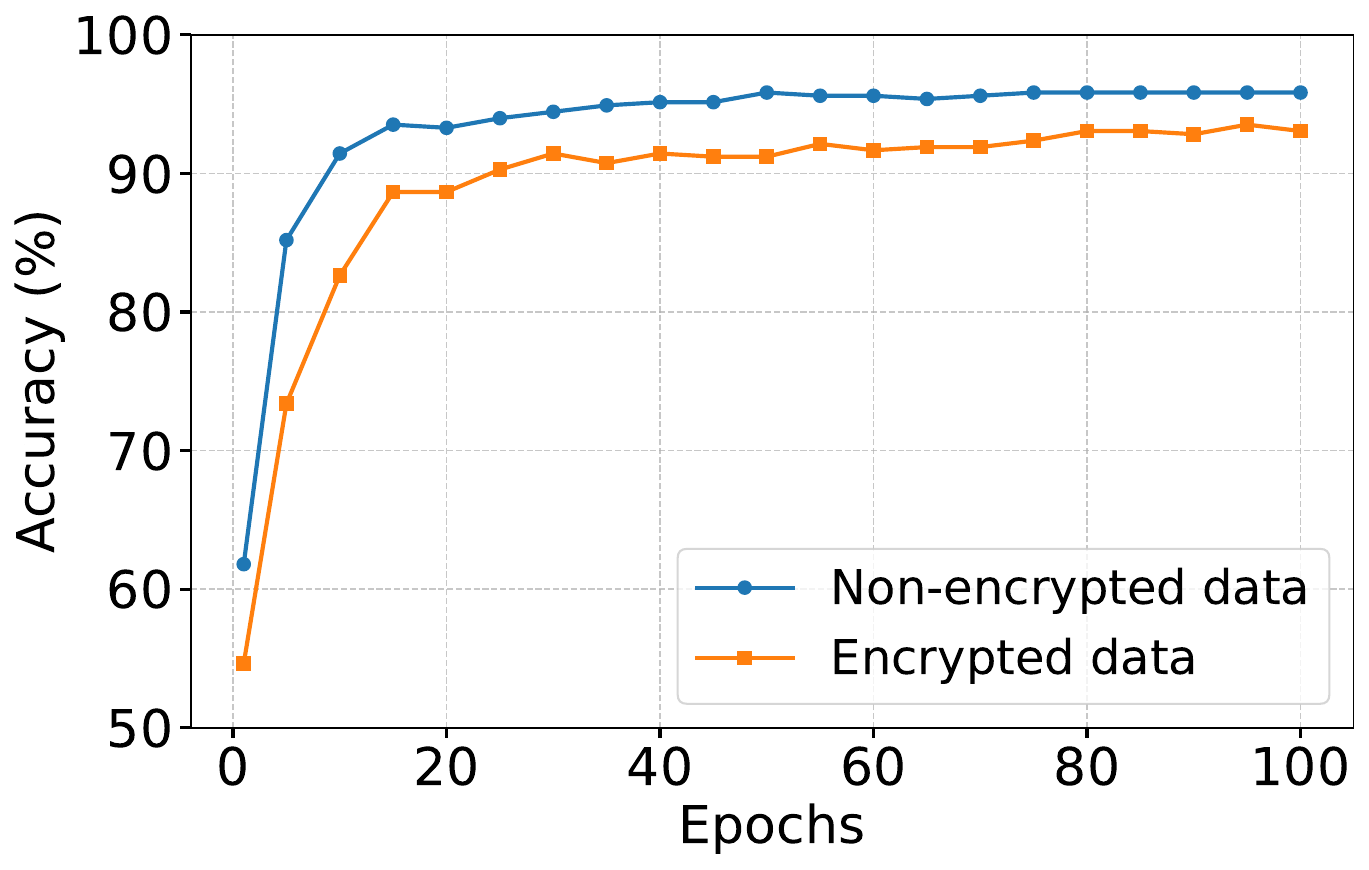}
        \caption{ReLU.}
        \label{fig:ReLU}
    \end{subfigure}
    % \hfill
    % Second subfigure
    \begin{subfigure}{0.35\textwidth}
        \includegraphics[width=\textwidth]{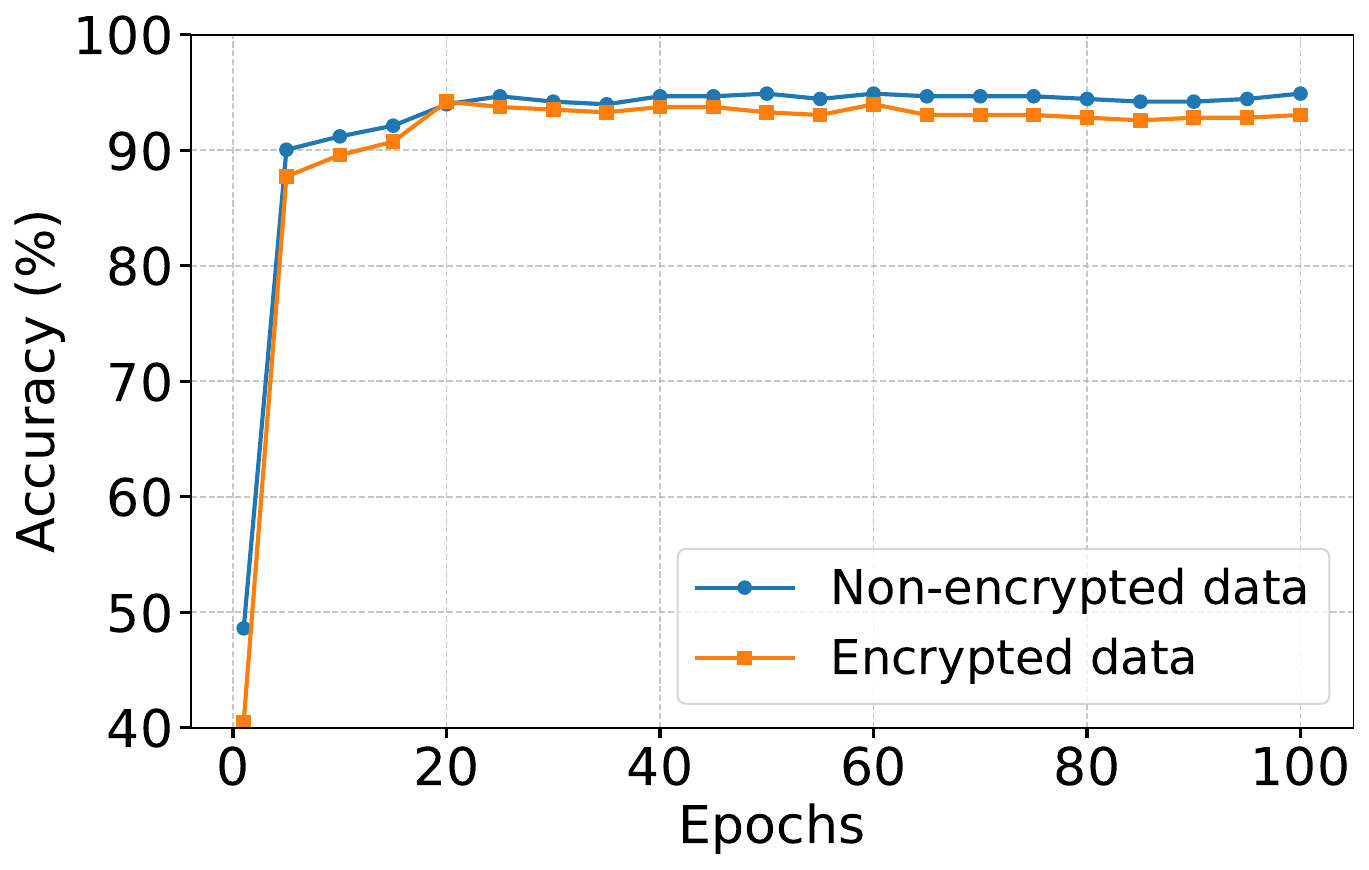}
        \caption{SiLU.}
        \label{fig:SiLU}
    \end{subfigure}
    % \hfill
    % Third subfigure
    \begin{subfigure}{0.35\textwidth}
        \includegraphics[width=\textwidth]{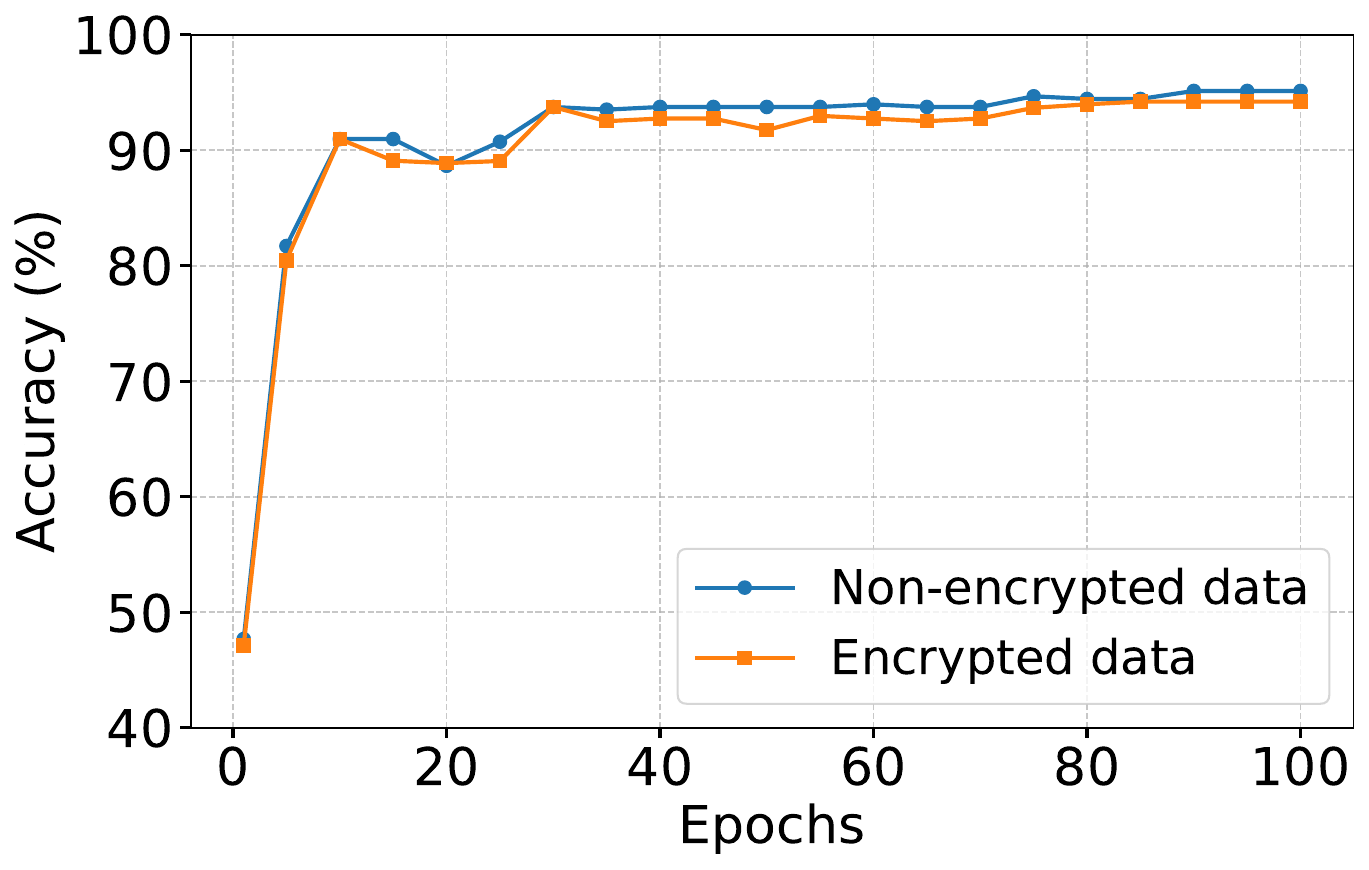}
        \caption{Square.}
        \label{fig:Square}
    \end{subfigure}
    \caption{Evolution of the model's accuracy over training epochs.}
    \label{fig:accuracy}
\end{figure}

\begin{figure}[htb!]
    \centering
    % First subfigure
    \begin{subfigure}{0.35\textwidth}
        \includegraphics[width=\textwidth]{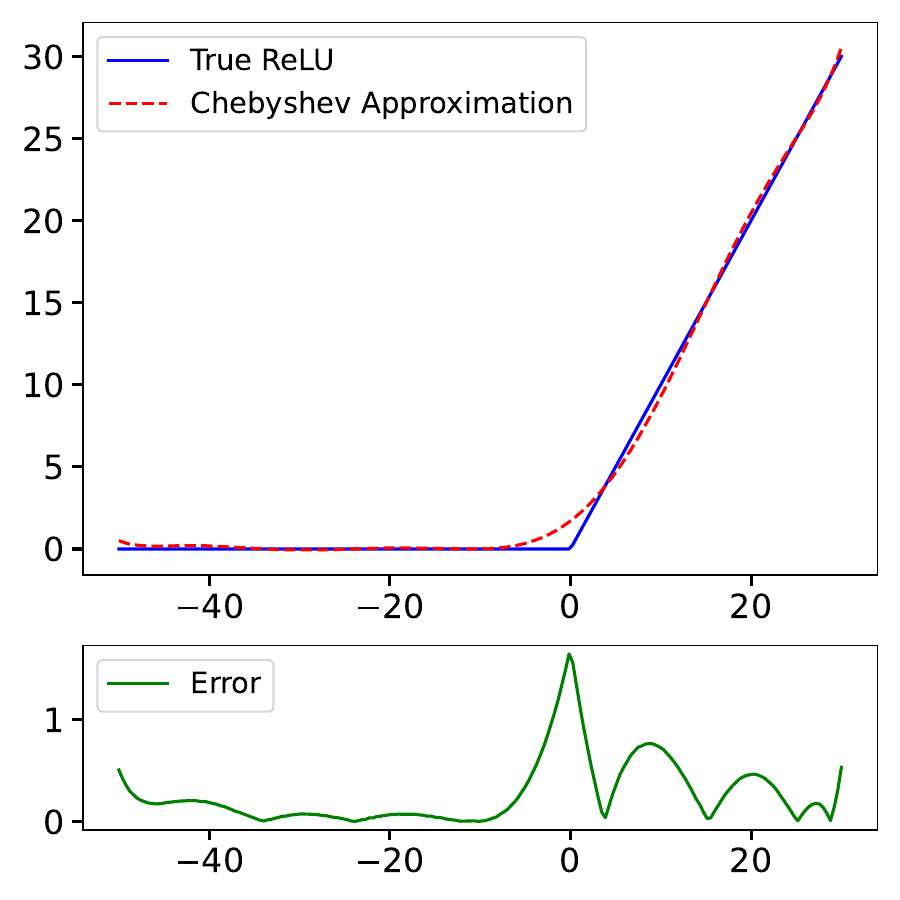}
        \caption{ReLU.}
        \label{fig:chebyshev_relu}
    \end{subfigure}
    \hfill
    % Second subfigure
    \begin{subfigure}{0.35\textwidth}
        \includegraphics[width=\textwidth]{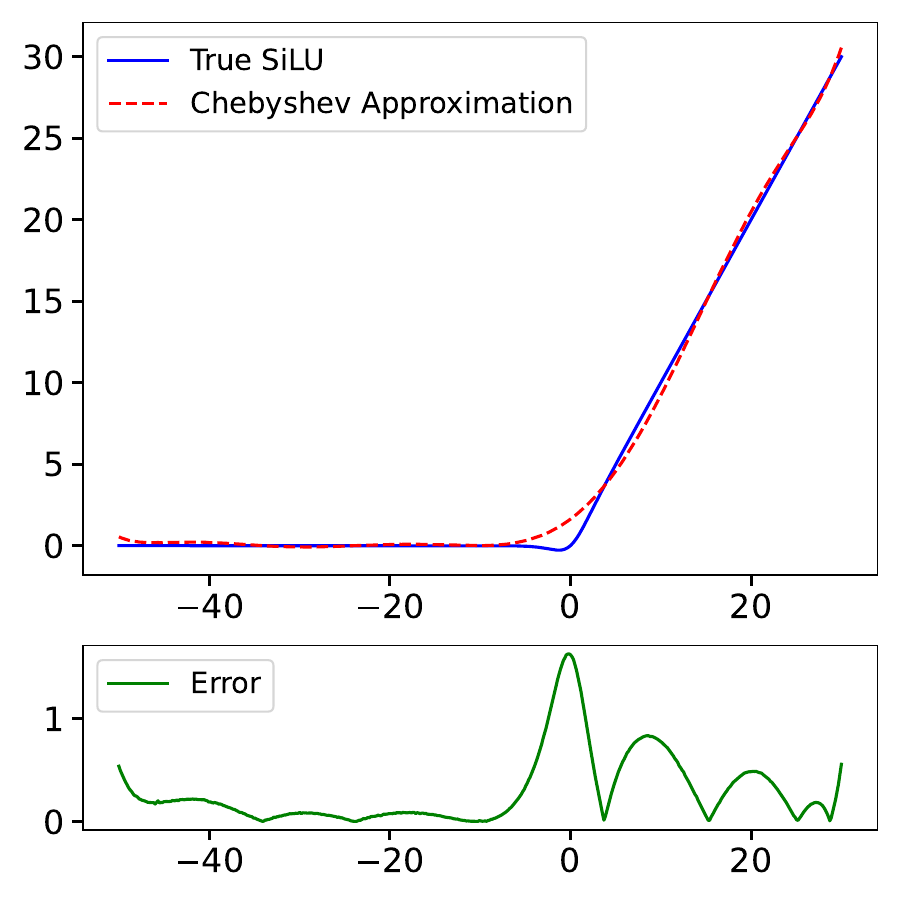}
        \caption{SiLU.}
        \label{fig:chebyshev_silu}
    \end{subfigure}
    \hfill
    % Third subfigure
    \begin{subfigure}{0.35\textwidth}
        \includegraphics[width=\textwidth]{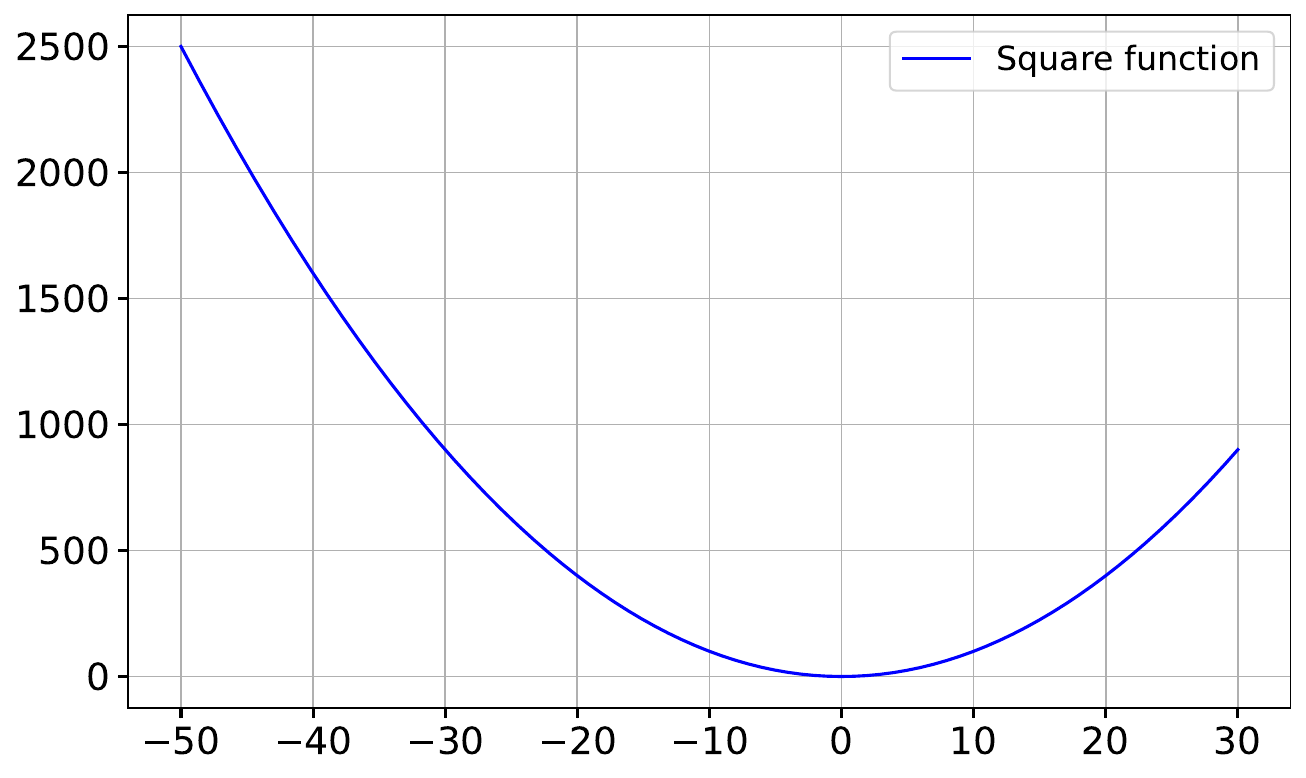}
        \caption{Square.}
        \label{fig:square}
    \end{subfigure}

    \caption{(a) Chebyshev approximation of ReLU, (b) Chebyshev approximation of SiLU, and (c) illustration of the Square function.}
    \label{fig:approximate}
\end{figure}

\subsubsection{Chebyshev Approximation of the Activation Functions}
\label{sec:D-II}
Figs.~\ref{fig:chebyshev_relu} and \ref{fig:chebyshev_silu} depict the Chebyshev approximation for the ReLU and SiLU activation functions. These figures reveal the significant differences in accuracy depending on the continuity characteristics of each function. For the SiLU function, the Chebyshev approximation performs more effectively due to the the smooth and continuous nature of the SiLU function, reducing errors across the entire value domain. In contrast, the Chebyshev approximation for the ReLU function works well in the region where \( x > 0 \) but encounters difficulties in the region where \( x < 0 \) due to the discontinuity at \( x = 0 \), leading to higher errors. On the other hand, the square function (\( x^2 \)) is simple and continuously smooth, thus not requiring the Chebyshev approximation, as shown in Fig.~\ref{fig:square}.

\subsubsection{Impact of the HE Security Level} 

\begin{table}[!] % * giúp bảng mở rộng toàn trang
\caption{Impact of LWE-based security level on the encrypted data classification accuracy.}
\label{tab:effect_poly_modulus_degree}
\centering
\resizebox{0.5\textwidth}{!}{ % Tăng kích thước bảng lên 120% chiều rộng
\begin{tabular}{cccc}
\toprule
\textbf{HE Security Level ($\boldsymbol{256}$-bit security)} 
% \textbf{Encryption's security level}
& \textbf{ReLU} & \textbf{SiLU} & \textbf{Square} \\ 
\cmidrule[0.4pt](lr{0.12em}){1-1}
\cmidrule[0.4pt](lr{0.12em}){2-2}
\cmidrule[0.4pt](lr{0.12em}){3-3}
\cmidrule[0.4pt](lr{0.12em}){4-4}
$8192$  & $\boldsymbol{95.139}\%$ & $94.676\%$ & $94.907\%$ \\ 
$16384$ & $94.907\%$ & $94.676\%$ & $\boldsymbol{95.139}\%$ \\ 
$32768$ & $93.056\%$ & $94.676\%$ & $\boldsymbol{95.139}\%$ \\ 
\bottomrule
\end{tabular}
}
\end{table}
% \textbf{revise}

Homomorphic Encryption (HE) is grounded in the Learning With Errors (LWE) problem, a post-quantum hardness assumption that underpins its robust security. A critical parameter influencing the security of HE schemes is the ring dimension, which directly affects the encryption strength and can achieve a $256$-bit security level when appropriately configured.

To balance the trade-off between security and computational efficiency, our experiments utilize polynomial modulus degrees of $8192$, $16384$, and $32768$. A polynomial modulus degree of $8192$ provides moderate security, suitable for specific applications, whereas degrees of $16384$ and $32768$ offer heightened security, meeting post-quantum cryptographic standards. Although increasing the polynomial modulus degree enhances encryption strength, it also incurs greater computational overhead, thereby impacting overall performance.

Table~\ref{tab:effect_poly_modulus_degree} illustrates the impact of the security level of encryption on the model's accuracy with encrypted data using ReLU, SiLU, and square activation functions. Specifically, for the ReLU activation function, the accuracy gradually decreases from $95.139\%$ at the level of $8192$ to $94.907\%$ at $16384$, and further down to $93.056\%$ at $32768$. This indicates that increasing the security level may degrade the performance of the model utilizing ReLU, likely due to ReLU being a non-differentiable function at zero, leading to significant errors at $x = 0$. In contrast, for the square function, accuracy remains stable and tends to increase from $94.907\%$ to $95.139\%$, demonstrating better adaptability of this function to homomorphic encryption. Meanwhile, the model using the SiLU function maintains a constant accuracy of $94.676\%$ across all encryption's security levels (with \textnormal{poly\_modulus\_degree} ranging from $8192$ to $32768$), indicating absolute stability of this function with increased encryption complexity.

\subsubsection{Performance with UAV-based Facial Images}

% This section presents the performance evaluation of our proposed model for UAV-based facial images, compared with the benchmark scheme (without encrypted data).

    This section evaluates the impact of varying HE security levels on the performance of our proposed model for UAV-based facial recognition, comparing it against the benchmark scheme without encryption.
    % This section evaluates the impact of varying HE security levels on the performance of our proposed model for UAV-based facial recognition, comparing it against the benchmark scheme without encryption.}
The testing results obtained after $300$ epochs of training are presented as performance metrics in Table~\ref{tab:metrics_comparison}, along with the model's confusion matrix shown in Figs.~\ref{fig:matrix_non} and~\ref{fig:matrix_enc}.

\begin{table}[htb]
\centering
\caption{Performance comparison between the proposed approach (PA) with different activation functions (Square, ReLU, and SiLU) and the benchmark scheme (without encrypted data).}
\label{tab:metrics_comparison}
% \resizebox{0.8\textwidth}{!}{
\begin{tabular}{lcccc}
\toprule
\textbf{Method} & \textbf{Accuracy} & \textbf{F1-score} & \textbf{Recall} & \textbf{Precision} \\ 
\midrule
% \cmidrule[0.4pt](lr{0.12em}){1-1}
% \cmidrule[0.4pt](lr{0.12em}){2-2}
% \cmidrule[0.4pt](lr{0.12em}){3-3}
% \cmidrule[0.4pt](lr{0.12em}){4-4}
% \cmidrule[0.4pt](lr{0.12em}){5-5}
PA w/ Square func.       
    & $94.676\%$
    & $0.9468$           
    & $0.9478$         
    & $0.9479$ \\ 
{PA w/  ReLU func.}         
    & $\boldsymbol{95.139\%}$                
    & $\boldsymbol{0.9502}$          
    & $\boldsymbol{0.9514}$          
    & $\boldsymbol{0.9499}$ \\ 
PA w/ SiLU func.        
    & $94.907\%$              
    & $0.9495$           
    & $0.9487$          
    & $0.9515$  \\ 
\midrule 
Benchmark        
    & $95.602\%$               
    & $0.9548$           
    & $0.9562$         
    & $0.9545$ \\ 
\bottomrule
\end{tabular}
% }
\end{table}

% \begin{figure}[!]
%     \centering
%     \begin{subfigure}[b]{0.5\textwidth}
%         \centering
%         \includegraphics[width=\linewidth]{Ảnh1.png}
%         \captionsetup{justification=raggedright, singlelinecheck=false, margin=2cm}
        
%         \label{fig:pre_model}
%     \end{subfigure}%

%     \caption{Confusion Matrix}
%     \label{fig:sys}
% \end{figure}

%Duc viet:
\noindent\textbf{HE security levels.}
We first examine the impact of different HE security levels (i.e., varying encryption ring sizes) on system performance in UAV-based face recognition tasks. The results indicate that at lower security levels ($8192$), the model achieves the highest or near-optimal accuracy, particularly with the ReLU activation function, which attains $95.139\%$, closely matching the $95.602\%$ accuracy of the non-encrypted model. This suggests that for UAV missions requiring real-time processing, such as search and rescue operations or crowd monitoring, selecting a moderate security level can help maintain performance while minimizing latency. Conversely, at the highest security level (32768), accuracy remains above $93\%$, but the substantial increase in computational cost may hinder responsiveness in emergency scenarios. Notably, the Square activation function maintains the highest accuracy at this security level ($95.139\%$), demonstrating stability and adaptability to deep ciphertext structures. As a result, UAV systems operating in environments with strict security requirements, such as border surveillance or critical infrastructure monitoring, may prioritize this configuration with the Square function, accepting computational trade-offs to ensure maximum privacy. These findings highlight the need for flexible HE security level selection, tailored to specific mission requirements and practical deployment conditions in UAV systems.

% We first examine the impact of different HE security levels—corresponding to varying encryption ring sizes—on system performance in UAV-based face recognition tasks. The results indicate that at lower security levels (8192), the model achieves the highest or near-optimal accuracy, particularly with the ReLU activation function, which attains 95.139$\%$, closely matching the 95.602$\%$ accuracy of the non-encrypted model. This suggests that for UAV missions requiring real-time processing, such as search-and-rescue operations or crowd monitoring, selecting a moderate security level can help maintain performance while minimizing latency. Conversely, at the highest security level (32768), accuracy remains above 93$\%$, but the substantial increase in computational cost may hinder responsiveness in emergency scenarios. Notably, the Square activation function maintains the highest accuracy at this security level (95.139$\%$), demonstrating stability and adaptability to deep ciphertext structures. As a result, UAV systems operating in environments with strict security requirements—such as border surveillance or critical infrastructure monitoring—may prioritize this configuration with the Square function, accepting computational trade-offs to ensure maximum privacy. These findings highlight the need for flexible HE security level selection, tailored to specific mission requirements and practical deployment conditions in UAV systems.
% }

\noindent\textbf{Accuracy.}
Table~\ref{tab:metrics_comparison} demonstrates that the model using non-encrypted data achieves the highest accuracy of $95.602\%$, indicating no performance degradation when data is unprotected. Conversely, the model using encrypted data with the square activation function achieves the lowest accuracy of $94.676\%$, likely due to limitations in nonlinear representation. The model using encrypted data with the SiLU activation function achieves a higher accuracy ($94.907\%$) than that with the square activation function ($94.676\%$), but lower than that when using ReLU ($95.139\%$). This may be attributed to the smooth nature of SiLU in reducing errors in encrypted data, although the presence of division and sigmoid functions usually leads to higher computational cost~\cite{silu_activation}. %\lqt{[Trung: Need a ref here?].}

The model using encrypted data with the ReLU activation function achieves an accuracy of $95.139\%$, close to that using the non-encrypted model, indicating that security measures do not significantly compromise accuracy. Although data encryption may result in a slight reduction in accuracy and other metrics, the encryption method utilizing ReLU shows significant potential by maintaining performance nearly equivalent to the non-encrypted model. This suggests that the ReLU-based encryption method can provide data security without substantially compromising detection and classification capabilities while ensuring high levels of accuracy, F1-score, recall, and precision. Therefore, the ReLU encryption model is a viable and effective choice for data security-aware applications.

%The model utilizing non-encrypted data achieved the highest accuracy at 95.602\%, indicating no performance degradation when data is unprotected. Conversely, the model using encrypted data with a square activation function attained the lowest accuracy at 94.676\%, potentially due to limitations in non-linear representation. The model employing encrypted data with a ReLU activation function achieved an accuracy of 95.139\%, closely approaching the non-encrypted model, suggesting that security measures do not significantly compromise accuracy.

%Although data encryption may lead to slight reductions in accuracy and other metrics, the encryption method using ReLU demonstrates significant potential by maintaining performance nearly equivalent to the non-encrypted model. This indicates that the ReLU-based encryption method can provide data security without substantially diminishing detection and classification capabilities, while still ensuring high levels of accuracy, F1-score, recall, and precision. Therefore, the ReLU encryption model is a viable and effective choice for applications requiring data security.

%Duc viet:
% The confusion matrix illustrates the high accuracy of the classification model, evidenced by the dominance of values along the diagonal. This pattern underscores the model's low misclassification rate, with relatively few entries off the diagonal. 

% Confusion matrix

\begin{figure}[htb]
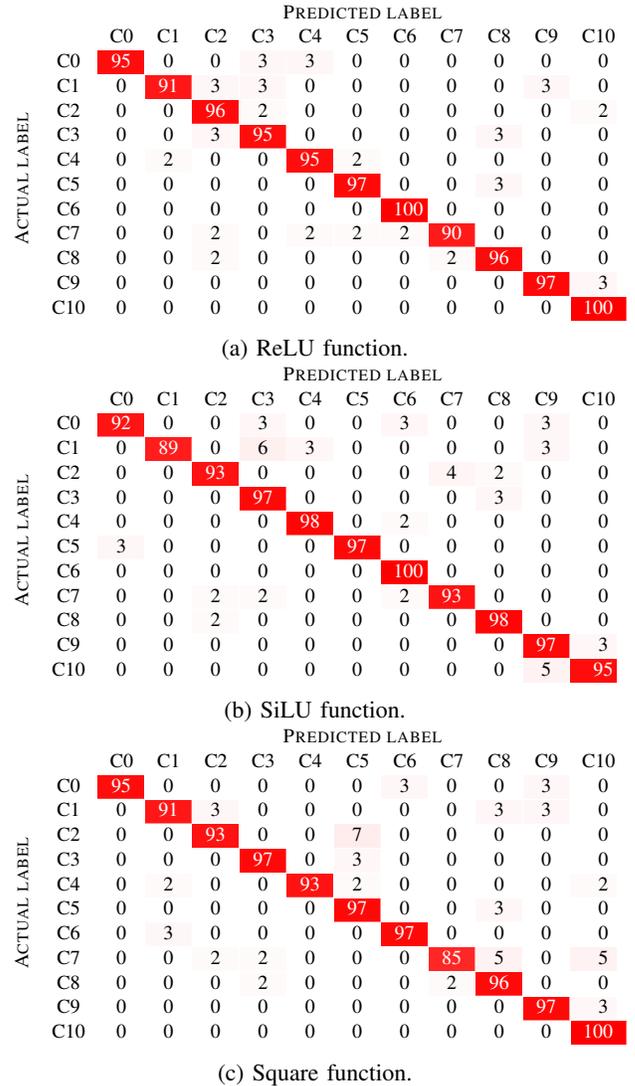

    \centering
    % First subfigure
    \begin{subfigure}{\columnwidth}
    \centering
        \adjustbox{max width=1.0\columnwidth}{% 
        {
            \newcommand\items{11}   % Number of classes
            \arrayrulecolor{white} % Table line colors
            
            \begin{tabular}{cc*{\items}{|E}|}
            \multicolumn{1}{c}{} 
            & \multicolumn{1}{c}{} 
            & \multicolumn{\items}{c}{\sc{}Predicted label} \\ 
            \hhline{~*\items{|-}|}
            \multicolumn{1}{c}{} & 
            \multicolumn{1}{c}{} & 
            \multicolumn{1}{c}{C0} & 
            \multicolumn{1}{c}{C1} & 
            \multicolumn{1}{c}{C2} &
            \multicolumn{1}{c}{C3} &
            \multicolumn{1}{c}{C4} &
            \multicolumn{1}{c}{C5} &
            \multicolumn{1}{c}{C6} &
            \multicolumn{1}{c}{C7} &
            \multicolumn{1}{c}{C8} &
            \multicolumn{1}{c}{C9} &
            \multicolumn{1}{c}{C10} \\ 
            \hhline{~*\items{|-}|}
            \multirow{\items}{*}{\rotatebox{90}{\sc{}Actual label}} 
            & C0 & 95 & 0 & 0 & 3 & 3 & 0 & 0 & 0 & 0 & 0 & 0\\ \hhline{~*\items{|-}|}
            & C1 & 0 & 91 & 3 & 3 & 0 & 0 & 0 & 0 & 0 & 3 & 0\\ \hhline{~*\items{|-}|}
            & C2 & 0 & 0 & 96 & 2 & 0 & 0 & 0 & 0 & 0 & 0 & 2\\ \hhline{~*\items{|-}|}
            & C3 & 0 & 0 & 3 & 95 & 0 & 0 & 0 & 0 & 3 & 0 & 0\\ \hhline{~*\items{|-}|}
            & C4 & 0 & 2 & 0 & 0 & 95 & 2 & 0 & 0 & 0 & 0 & 0\\ \hhline{~*\items{|-}|}
            & C5 & 0 & 0 & 0 & 0 & 0 & 97 & 0 & 0 & 3 & 0 & 0\\ \hhline{~*\items{|-}|}
            & C6 & 0 & 0 & 0 & 0 & 0 & 0 & 100 & 0 & 0 & 0 & 0\\ \hhline{~*\items{|-}|}
            & C7 & 0 & 0 & 2 & 0 & 2 & 2 & 2 & 90 & 0 & 0 & 0\\ \hhline{~*\items{|-}|}
            & C8 & 0 & 0 & 2 & 0 & 0 & 0 & 0 & 2 & 96 & 0 & 0\\ \hhline{~*\items{|-}|}
            & C9 & 0 & 0 & 0 & 0 & 0 & 0 & 0 & 0 & 0 & 97 & 3\\ \hhline{~*\items{|-}|}
            & C10 & 0 & 0 & 0 & 0 & 0 & 0 & 0 & 0 & 0 & 0 & \multicolumn{1}{|>{\columncolor{red!100}}l|}{\textcolor{white}{100}}\\ \hhline{~*\items{|-}|}

            \end{tabular}
            \arrayrulecolor{black} % reset table line color
        }
        }
        \caption{ReLU function.}
        \label{fig:confusion_ReLu}
    \end{subfigure}
    \hfill
    % Second subfigure
    \begin{subfigure}{\columnwidth}
        \centering
        \adjustbox{max width=1.0\columnwidth}{% 
        {
            \newcommand\items{11}   % Number of classes
            \arrayrulecolor{white} % Table line colors
            \begin{tabular}{cc*{\items}{|E}|}
            \multicolumn{1}{c}{} 
            & \multicolumn{1}{c}{} 
            & \multicolumn{\items}{c}{\sc{}Predicted label} \\ 
            \hhline{~*\items{|-}|}
            \multicolumn{1}{c}{} & 
            \multicolumn{1}{c}{} & 
            \multicolumn{1}{c}{C0} & 
            \multicolumn{1}{c}{C1} & 
            \multicolumn{1}{c}{C2} &
            \multicolumn{1}{c}{C3} &
            \multicolumn{1}{c}{C4} &
            \multicolumn{1}{c}{C5} &
            \multicolumn{1}{c}{C6} &
            \multicolumn{1}{c}{C7} &
            \multicolumn{1}{c}{C8} &
            \multicolumn{1}{c}{C9} &
            \multicolumn{1}{c}{C10} \\ 
            \hhline{~*\items{|-}|}
            \multirow{\items}{*}{\rotatebox{90}{\sc{}Actual label}} 
            & C0 & 92 & 0 & 0 & 3 & 0 & 0 & 3 & 0 & 0 & 3 & 0\\ \hhline{~*\items{|-}|}
            & C1 & 0 & 89 & 0 & 6 & 3 & 0 & 0 & 0 & 0 & 3 & 0\\ \hhline{~*\items{|-}|}
            & C2 & 0 & 0 & 93 & 0 & 0 & 0 & 0 & 4 & 2 & 0 & 0\\ \hhline{~*\items{|-}|}
            & C3 & 0 & 0 & 0 & 97 & 0 & 0 & 0 & 0 & 3 & 0 & 0\\ \hhline{~*\items{|-}|}
            & C4 & 0 & 0 & 0 & 0 & 98 & 0 & 2 & 0 & 0 & 0 & 0\\ \hhline{~*\items{|-}|}
            & C5 & 3 & 0 & 0 & 0 & 0 & 97 & 0 & 0 & 0 & 0 & 0\\ \hhline{~*\items{|-}|}
            & C6 & 0 & 0 & 0 & 0 & 0 & 0 & 100 & 0 & 0 & 0 & 0\\ \hhline{~*\items{|-}|}
            & C7 & 0 & 0 & 2 & 2 & 0 & 0 & 2 & 93 & 0 & 0 & 0\\ \hhline{~*\items{|-}|}
            & C8 & 0 & 0 & 2 & 0 & 0 & 0 & 0 & 0 & 98 & 0 & 0\\ \hhline{~*\items{|-}|}
            & C9 & 0 & 0 & 0 & 0 & 0 & 0 & 0 & 0 & 0 & 97 & 3\\ \hhline{~*\items{|-}|}
            & C10 & 0 & 0 & 0 & 0 & 0 & 0 & 0 & 0 & 0 & 5 & 95\\ \hhline{~*\items{|-}|}

            \end{tabular}
            \arrayrulecolor{black} % reset table line color
        }
        }
        \caption{SiLU function.}
        \label{fig:confusion_SiLU}
    \end{subfigure}
    \hfill
    % Third subfigure
    \begin{subfigure}{\columnwidth}
        \centering
        \adjustbox{max width=1.0\columnwidth}{% 
        {
            \newcommand\items{11}   % Number of classes
            \arrayrulecolor{white} % Table line colors
            \begin{tabular}{cc*{\items}{|E}|}
            \multicolumn{1}{c}{} 
            & \multicolumn{1}{c}{} 
            & \multicolumn{\items}{c}{\sc{}Predicted label} \\ 
            \hhline{~*\items{|-}|}
            \multicolumn{1}{c}{} & 
            \multicolumn{1}{c}{} & 
            \multicolumn{1}{c}{C0} & 
            \multicolumn{1}{c}{C1} & 
            \multicolumn{1}{c}{C2} &
            \multicolumn{1}{c}{C3} &
            \multicolumn{1}{c}{C4} &
            \multicolumn{1}{c}{C5} &
            \multicolumn{1}{c}{C6} &
            \multicolumn{1}{c}{C7} &
            \multicolumn{1}{c}{C8} &
            \multicolumn{1}{c}{C9} &
            \multicolumn{1}{c}{C10} \\ 
            \hhline{~*\items{|-}|}
            \multirow{\items}{*}{\rotatebox{90}{\sc{}Actual label}} 
            & C0 & 95 & 0 & 0 & 0 & 0 & 0 & 3 & 0 & 0 & 3 & 0 \\ \hhline{~*\items{|-}|}
            & C1 & 0 & 91 & 3 & 0 & 0 & 0 & 0 & 0 & 3 & 3 & 0 \\ \hhline{~*\items{|-}|}
            & C2 & 0 & 0 & 93 & 0 & 0 & 7 & 0 & 0 & 0 & 0 & 0 \\ \hhline{~*\items{|-}|}
            & C3 & 0 & 0 & 0 & 97 & 0 & 3 & 0 & 0 & 0 & 0 & 0 \\ \hhline{~*\items{|-}|}
            & C4 & 0 & 2 & 0 & 0 & 93 & 2 & 0 & 0 & 0 & 0 & 2 \\ \hhline{~*\items{|-}|}
            & C5 & 0 & 0 & 0 & 0 & 0 & 97 & 0 & 0 & 3 & 0 & 0 \\ \hhline{~*\items{|-}|}
            & C6 & 0 & 3 & 0 & 0 & 0 & 0 & 97 & 0 & 0 & 0 & 0 \\ \hhline{~*\items{|-}|}
            & C7 & 0 & 0 & 2 & 2 & 0 & 0 & 0 & 85 & 5 & 0 & 5 \\ \hhline{~*\items{|-}|}
            & C8 & 0 & 0 & 0 & 2 & 0 & 0 & 0 & 2 & 96 & 0 & 0 \\ \hhline{~*\items{|-}|}
            & C9 & 0 & 0 & 0 & 0 & 0 & 0 & 0 & 0 & 0 & 97 & 3 \\ \hhline{~*\items{|-}|}
            & C10 & 0 & 0 & 0 & 0 & 0 & 0 & 0 & 0 & 0 & 0 & \multicolumn{1}{|>{\columncolor{red!100}}l|}{\textcolor{white}{100}}\\ \hhline{~*\items{|-}|}
            \end{tabular}
            \arrayrulecolor{black} % reset table line color
        }
        }
        \caption{Square function.}
        \label{fig:confusion_Square}
    \end{subfigure}

    \caption{Confusion matrix (elements shown in $\%$) of the proposed method for non-encrypted data using (a) ReLU, (b) SiLU, and (c) Square function.}
    \label{fig:matrix_non}
\end{figure}

%enc
\begin{figure}[htb]
    \centering
    % First subfigure
    \begin{subfigure}{\columnwidth}
        \centering
        \adjustbox{max width=1.0\columnwidth}{% 
        {
            \newcommand\items{11}   % Number of classes
            \arrayrulecolor{white} % Table line colors
            \begin{tabular}{cc*{\items}{|E}|}
            \multicolumn{1}{c}{} 
            & \multicolumn{1}{c}{} 
            & \multicolumn{\items}{c}{\sc{}Predicted label} \\ 
            \hhline{~*\items{|-}|}
            \multicolumn{1}{c}{} & 
            \multicolumn{1}{c}{} & 
            \multicolumn{1}{c}{C0} & 
            \multicolumn{1}{c}{C1} & 
            \multicolumn{1}{c}{C2} &
            \multicolumn{1}{c}{C3} &
            \multicolumn{1}{c}{C4} &
            \multicolumn{1}{c}{C5} &
            \multicolumn{1}{c}{C6} &
            \multicolumn{1}{c}{C7} &
            \multicolumn{1}{c}{C8} &
            \multicolumn{1}{c}{C9} &
            \multicolumn{1}{c}{C10} \\ 
            \hhline{~*\items{|-}|}
            \multirow{\items}{*}{\rotatebox{90}{\sc{}Actual label}} 
            & C0 & 95 & 0 & 0 & 3 & 0 & 0 & 3 & 0 & 0 & 0 & 0\\ \hhline{~*\items{|-}|}
            & C1 & 0 & 91 & 3 & 3 & 0 & 0 & 0 & 0 & 0 & 3 & 0\\ \hhline{~*\items{|-}|}
            & C2 & 0 & 0 & 96 & 2 & 0 & 0 & 0 & 0 & 0 & 0 & 2\\ \hhline{~*\items{|-}|}
            & C3 & 0 & 0 & 0 & 95 & 0 & 0 & 3 & 0 & 3 & 0 & 0\\ \hhline{~*\items{|-}|}
            & C4 & 0 & 2 & 0 & 0 & 98 & 0 & 0 & 0 & 0 & 0 & 0\\ \hhline{~*\items{|-}|}
            & C5 & 0 & 0 & 0 & 0 & 0 & 97 & 0 & 0 & 3 & 0 & 0\\ \hhline{~*\items{|-}|}
            & C6 & 0 & 0 & 0 & 0 & 0 & 0 & 100 & 0 & 0 & 0 & 0\\ \hhline{~*\items{|-}|}
            & C7 & 0 & 0 & 0 & 0 & 2 & 2 & 2 & 90 & 0 & 2 & 0\\ \hhline{~*\items{|-}|}
            & C8 & 0 & 0 & 2 & 0 & 0 & 2 & 0 & 2 & 94 & 0 & 0\\ \hhline{~*\items{|-}|}
            & C9 & 0 & 0 & 0 & 0 & 0 & 0 & 0 & 0 & 0 & 94 & 6\\ \hhline{~*\items{|-}|}
            & C10 & 0 & 0 & 0 & 0 & 0 & 0 & 3 & 0 & 0 & 0 & 97\\ \hhline{~*\items{|-}|}

            \end{tabular}
            \arrayrulecolor{black} % reset table line color
        }
        }
        \caption{ReLU function.}
        \label{fig:confusion_enc_ReLu}
    \end{subfigure}
    \hfill
    % Second subfigure
    \begin{subfigure}{\columnwidth}
        \centering
        \adjustbox{max width=1.0\columnwidth}{% 
        {
            \newcommand\items{11}   % Number of classes
            \arrayrulecolor{white} % Table line colors
            \begin{tabular}{cc*{\items}{|E}|}
            \multicolumn{1}{c}{} 
            & \multicolumn{1}{c}{} 
            & \multicolumn{\items}{c}{\sc{}Predicted label} \\ 
            \hhline{~*\items{|-}|}
            \multicolumn{1}{c}{} & 
            \multicolumn{1}{c}{} & 
            \multicolumn{1}{c}{C0} & 
            \multicolumn{1}{c}{C1} & 
            \multicolumn{1}{c}{C2} &
            \multicolumn{1}{c}{C3} &
            \multicolumn{1}{c}{C4} &
            \multicolumn{1}{c}{C5} &
            \multicolumn{1}{c}{C6} &
            \multicolumn{1}{c}{C7} &
            \multicolumn{1}{c}{C8} &
            \multicolumn{1}{c}{C9} &
            \multicolumn{1}{c}{C10} \\ 
            \hhline{~*\items{|-}|}
            \multirow{\items}{*}{\rotatebox{90}{\sc{}Actual label}} 
            & C0 & 92 & 0 & 0 & 3 & 0 & 0 & 3 & 0 & 0 & 3 & 0\\ \hhline{~*\items{|-}|}
            & C1 & 0 & 86 & 0 & 3 & 6 & 0 & 3 & 0 & 0 & 3 & 0\\ \hhline{~*\items{|-}|}
            & C2 & 0 & 0 & 96 & 0 & 0 & 0 & 0 & 2 & 2 & 0 & 0\\ \hhline{~*\items{|-}|}
            & C3 & 0 & 0 & 0 & 97 & 0 & 0 & 0 & 0 & 3 & 0 & 0\\ \hhline{~*\items{|-}|}
            & C4 & 0 & 0 & 0 & 0 & 100 & 0 & 0 & 0 & 0 & 0 & 0\\ \hhline{~*\items{|-}|}
            & C5 & 3 & 0 & 0 & 0 & 0 & 97 & 0 & 0 & 0 & 0 & 0\\ \hhline{~*\items{|-}|}
            & C6 & 0 & 0 & 0 & 0 & 5 & 0 & 95 & 0 & 0 & 0 & 0\\ \hhline{~*\items{|-}|}
            & C7 & 0 & 0 & 2 & 2 & 0 & 5 & 2 & 83 & 5 & 0 & 0\\ \hhline{~*\items{|-}|}
            & C8 & 0 & 0 & 2 & 0 & 2 & 0 & 0 & 0 & 96 & 0 & 0\\ \hhline{~*\items{|-}|}
            & C9 & 0 & 0 & 3 & 0 & 0 & 0 & 0 & 0 & 0 & 97 & 0\\ \hhline{~*\items{|-}|}
            & C10 & 0 & 0 & 3 & 0 & 0 & 0 & 3 & 0 & 0 & 5 & 90\\ \hhline{~*\items{|-}|}

            \end{tabular}
            \arrayrulecolor{black} % reset table line color
        }
        }
        \caption{SiLU function.}
        \label{fig:confusion_enc_SiLU}
    \end{subfigure}
    \hfill
    % Third subfigure
    \begin{subfigure}{\columnwidth}
        \centering
        \adjustbox{max width=1.0\columnwidth}{% 
        {
            \newcommand\items{11}   % Number of classes
            \arrayrulecolor{white} % Table line colors
            \begin{tabular}{cc*{\items}{|E}|}
            \multicolumn{1}{c}{} 
            & \multicolumn{1}{c}{} 
            & \multicolumn{\items}{c}{\sc{}Predicted label} \\ 
            \hhline{~*\items{|-}|}
            \multicolumn{1}{c}{} & 
            \multicolumn{1}{c}{} & 
            \multicolumn{1}{c}{C0} & 
            \multicolumn{1}{c}{C1} & 
            \multicolumn{1}{c}{C2} &
            \multicolumn{1}{c}{C3} &
            \multicolumn{1}{c}{C4} &
            \multicolumn{1}{c}{C5} &
            \multicolumn{1}{c}{C6} &
            \multicolumn{1}{c}{C7} &
            \multicolumn{1}{c}{C8} &
            \multicolumn{1}{c}{C9} &
            \multicolumn{1}{c}{C10} \\ 
            \hhline{~*\items{|-}|}
            \multirow{\items}{*}{\rotatebox{90}{\sc{}Actual label}} 
            & C0 & 95 & 0 & 0 & 0 & 0 & 0 & 3 & 0 & 0 & 3 & 0\\ \hhline{~*\items{|-}|}
            & C1 & 0 & 91 & 3 & 0 & 0 & 0 & 0 & 0 & 3 & 3 & 0\\ \hhline{~*\items{|-}|}
            & C2 & 0 & 0 & 93 & 0 & 0 & 7 & 0 & 0 & 0 & 0 & 0\\ \hhline{~*\items{|-}|}
            & C3 & 0 & 0 & 0 & 97 & 0 & 3 & 0 & 0 & 0 & 0 & 0\\ \hhline{~*\items{|-}|}
            & C4 & 0 & 2 & 0 & 0 & 93 & 2 & 0 & 0 & 0 & 0 & 2\\ \hhline{~*\items{|-}|}
            & C5 & 0 & 0 & 0 & 0 & 0 & 97 & 0 & 0 & 3 & 0 & 0\\ \hhline{~*\items{|-}|}
            & C6 & 0 & 3 & 0 & 0 & 0 & 0 & 97 & 0 & 0 & 0 & 0\\ \hhline{~*\items{|-}|}
            & C7 & 0 & 0 & 2 & 2 & 0 & 0 & 0 & 85 & 5 & 0 & 5\\ \hhline{~*\items{|-}|}
            & C8 & 0 & 0 & 0 & 2 & 0 & 0 & 0 & 2 & 96 & 0 & 0\\ \hhline{~*\items{|-}|}
            & C9 & 0 & 0 & 0 & 0 & 0 & 0 & 0 & 0 & 0 & 97 & 3\\ \hhline{~*\items{|-}|}
            & C10 & 0 & 0 & 0 & 0 & 0 & 0 & 0 & 0 & 0 & 0 & \multicolumn{1}{|>{\columncolor{red!100}}l|}{\textcolor{white}{100}}\\ \hhline{~*\items{|-}|}

            \end{tabular}
            \arrayrulecolor{black} % reset table line color
        }
        }
        \caption{Square function.}
        \label{fig:confusion_enc_Square}
    \end{subfigure}

    \caption{Confusion matrix (elements shown in $\%$) of the proposed method for encrypted data using (a) ReLU, (b) SiLU, and (c) Square function.}
    \label{fig:matrix_enc}
\end{figure}

\vspace{0.1cm}
\noindent\textbf{Confusion matrix.}
Fig.~\ref{fig:matrix_non} depicts the confusion matrices for the models using ReLU, SiLU, and square activation functions with non-encrypted data. The confusion matrices for models using ReLU and SiLU activation functions show strong classification performance with high overall accuracy. Most classes exhibit low misclassification rates and superior accuracy, although there is still minor confusion in a few classes. For the square activation function, the misclassification rate is higher, resulting in slightly lower performance.

Fig.~\ref{fig:matrix_enc} depicts the confusion matrices when performing with encrypted data. The performance of all models decreases compared to non-encrypted data; however, they still maintain high accuracy in easily recognizable classes. Among them, the confusion  matrix for the model using the Square function is most affected by encryption, with a marked decline in performance. ReLU-based and SiLU-based models, on the other hand, maintain stable performance, with their confusion matrices showing no significant impact from encryption.

\section{Conclusion}
\label{sec:Conclusion}

In the paper, we have proposed a privacy-preserving machine learning approach for UAV-based face detection using Homomorphic Encryption (HE). By integrating HE with advanced neural networks, this method ensures that facial data remains secure with minimal impact on detection accuracy. Interestingly, although classifying directly on the encrypted data, our approach still can achieve a very high accuracy (with a gap of less than $0.1\%$ compared with the benchmark classifying on non-encrypted data). In addition, in experiments on the real dataset, we reveal that the model using HE with the ReLU activation function achieved results nearly equivalent to the performance of non-encrypted data models. These findings highlight the potential of HE-based methods to enhance the privacy and effectiveness of UAV-based surveillance systems, paving the way for future research and development in secure aerial face detection technologies.

Despite its promising contributions, this study has certain limitations. Specifically, since the primary objective of this work is to demonstrate the feasibility of deep learning inference on UAV image data encrypted using Homomorphic Encryption (HE), the experiments primarily focus on verifying the correctness of encrypted convolutional and fully connected layers, assessing the ability to maintain accuracy compared to unencrypted models, and evaluating computational feasibility under different CKKS security levels. Evaluation of performance metrics such as latency, throughput, and energy consumption on edge hardware during actual deployments has not yet been conducted within this study. This is because such an assessment requires a separate implementation effort, such as compiling code for ARM architectures or analyzing performance on microservers with GPU acceleration, which goes beyond the initial feasibility analysis presented in this paper.

Future research will focus on deploying the optimized CNN model alongside HE-based computations on practical edge computing environments. Additionally, we will design experiments to measure latency during the inference process across different stages, including encryption, data transmission, inference on encrypted data, and decryption, as well as analyze and optimize the system based on key performance factors such as processing speed, energy consumption, and scalability for UAV applications in operational environments. Establishing functional correctness and model accuracy is considered a fundamental first step, paving the way for technical advancements in fully encrypted inference workflows within UAV-edge ecosystems. These research directions hold significant potential for developing a comprehensive UAV-edge system while enabling new applications in secure mobile data processing and surveillance.

%\section*{Acknowledgment}
\bibliographystyle{IEEEtran}
\bibliography{ref}

\end{document}